\documentclass[aps,prb,twocolumn,reprint,noshowpacs,superscriptaddress,citeautoscript,longbibliography]{revtex4-1}
\usepackage{graphicx}
\usepackage{amsmath,amssymb}
\usepackage{fixmath}

\bibpunct{[}{]}{,}{n}{}{}

\usepackage{here} 

\DeclareMathOperator{\sgn}{sgn}

\begin{document}

\title{Topology and zero energy edge states in carbon nanotubes with
  superconducting pairing}

\author{W. Izumida}
\email[]{wizumida@tohoku.ac.jp}
\affiliation{Department of Physics, Tohoku University, Sendai 980-8578, Japan}
\affiliation{Institute of Theoretical Physics, University of Regensburg, 93040 Regensburg, Germany}
\author{L. Milz}
\affiliation{Institute of Theoretical Physics, University of Regensburg, 93040 Regensburg, Germany}
\author{M. Marganska}
\affiliation{Institute of Theoretical Physics, University of Regensburg, 93040 Regensburg, Germany}
\author{M. Grifoni}
\affiliation{Institute of Theoretical Physics, University of Regensburg, 93040 Regensburg, Germany}

\date{
}

\begin{abstract}
We investigate the spectrum of finite-length carbon nanotubes in the
presence of onsite and nearest-neighbor superconducting pairing terms.
A one-dimensional ladder-type lattice model is developed to explore
the low-energy spectrum and the nature of the electronic states.
We find that zero energy edge states can emerge in zigzag class carbon
nanotubes as a combined effect of curvature-induced Dirac point shift
and strong superconducting coupling between nearest-neighbor sites.
The chiral symmetry of the system is exploited to define a winding
number topological invariant.
The associated topological phase diagram shows regions with
nontrivial winding number in the plane of chemical potential and
superconducting nearest-neighbor pair potential (relative to the
onsite pair potential).
A one-dimensional continuum model reveals the topological origin of
the zero energy edge states:
A bulk-edge correspondence is proven, which shows that the condition
for nontrivial winding number and that for the emergence of edge
states are identical.
For armchair class nanotubes, the presence of edge states 
in the superconducting gap
depends on
the nanotube's boundary shape.
For the minimal boundary condition, the emergence of
the subgap states can also be deduced from the winding number.
\end{abstract}


\maketitle

\section{Introduction}

Single-wall carbon nanotubes (SWNTs) are one-dimensional (1D) crystals
where the graphene honeycomb lattice, with its pseudospin valley
degree of freedom, is rolled into a seamless cylinder.
The finite curvature of the nanotube surface combined with the
presence of valley and spin degrees of freedom is at the origin of a
large variety of peculiar quantum transport properties, which have
been intensively investigated in the last
decades~\cite{Laird-RevModPhys-2015}.
In recent studies, the emphasis has been put on the bound-state
spectrum which naturally arises due to the finiteness of the SWNT
length.
It has been shown that the valley degeneracy of the bound states is
not only lifted by the curvature-induced spin-orbit
interaction~\cite{Ando-2000-06,Chico-2004-10,Huertas-Hernando-2006-10,Kuemmeth-2008-03,Chico-2009-06,Izumida-2009-06,Jeong-2009-08,Jhang-2010-07,Jespersen-2011-04,Klinovaja-PRB-2011-08,delValle-PRB-2011-11,Steele-NatCommun-2013},
but also by a valley mixing from the
edges~\cite{Izumida-2015-06,Marganska-PRB-2015}.
Furthermore, open-ended SWNTs commonly host edge states whose energies
lie in the bulk band
gap~\cite{Sasaki-PRB-2005,Marganska-PRB-2011,Izumida-2015-06}.
Topological considerations can give a new perspective on the nature of
these localized states.
Recently, a one-to-one correspondence has been shown between the
number of edge states and a winding number topological
invariant~\cite{Izumida-2016-05}, and that a topological phase
transition can be induced by an external magnetic
field~\cite{Okuyama-2017-01}.
Although the topological argument does not give a detailed information
on the edge states (e.g., on their decay length), the use of
topological invariants enables a general discussion on the emergence
of the edge states, which is possible as long as the corresponding
bulk system keeps the band gap.

When a superconductor is connected to a normal conductor,
superconducting correlations leak into the normal
conductor~\cite{Tinkham-2004} and can give rise to a proximity induced
superconducting gap.
In confined nanoconductors such as quantum dots and
wires~\cite{Franceschi-2010-10}, resonant Andreev processes at the
superconductor--normal-metal interface cause the formation of bound
states with excitation energies below the superconducting gap,
referred to as Andreev bound states.
Such bound states have also been observed in SWNT-superconductor
hybrid
devices~\cite{Pillet-NP-2010,Kim-PRL-2013,Pillet-PRB-2013,Schindele-PRB-2014-01,Kumar-PRB-2014,Gramich-PRL-2015}.
Reflecting superconducting correlations, the bound states correspond
to entangled time-reversed electron-hole pairs and, hence, always come
in pairs of opposite energy with respect to the center of the gap.
Because the energy of the bound states depends on the microscopic
details of the nanoconductor, in some systems it is possible to induce
a crossing of the pair at zero energy upon variation of a gate voltage
or of an external magnetic
field~\cite{Pillet-NP-2010,Pillet-PRB-2013,Kumar-PRB-2014}.
Such states may like to stick at zero energy like a topological state,
as pointed out in recent works on superconducting
nanowires~\cite{Deng-Science-2016,Liu-cond-mat-2017-05}.
In this context it is interesting to have the possibility to
discriminate between nontopological bound states sticking at zero
energy and truly topological zero energy bound
states~\cite{Sato-RPP-2017}.

In this paper, we address theoretically the topological origin of zero
energy bound states localized at the edges of a SWNT proximity coupled
to a superconductor.
On the one hand, we perform numerical calculations of the spectrum of
SWNTs with length of a few micrometers which show that zero energy
edge states emerge in some regions of chemical potential and proximity
pairing strengths.
These calculations are based on a 1D lattice model which includes the
effects of curvature and superconductivity, and uses the
helical-angular symmetry of the system~\cite{White-1993-03}.
It extends the 1D lattice model of
Refs.~\cite{Izumida-2015-06,Izumida-2016-05,Okuyama-2017-01} to
the superconducting case.
On the other hand, the chiral symmetry of the bulk Hamiltonian allows
us to introduce a winding number as a topological invariant.
We show that the edge states emerge in the parameter region of
nontrivial, that is nonzero, winding number.
The condition for the nontrivial winding number will be given in
Eq.~\eqref{eq:cond_nontrivial_w} [and
  Eq.~\eqref{eq:cond_nontrivial_w_expr}].
The nontrivial winding number is the combined result of the
curvature-induced shift of the Dirac points from the $K$ or $K'$
points, and strong superconducting coupling between nearest neighbors.
Finally, a 1D continuum model is introduced which allows us to obtain
the condition for the emergence of the edge states, which will be
given in Eq.~\eqref{eq:cond_edgeState}.
By comparing with the previously obtained condition for a nontrivial
winding number, we find that these conditions are identical, hence
proving the bulk-edge correspondence in our system.
We notice that the formation of edge states depends not only on the
chemical potential and the pairing potentials, but also on the
chirality and the boundary shape of the nanotubes since they strongly
affect the coupling of the two valleys.

Since the zero energy bound states appear in the induced
superconducting gap region, these states can be regarded as Andreev
bound states.
In contrast to the conventional Andreev bound states, which extend in
the whole of the nanoconductor, the zero energy bound states we
observe are more specifically regarded as surface Andreev bound
states, which in our case are also of topological
origin~\cite{Kashiwaya-RPP-2000,Sato-RPP-2017}.

\begin{figure}[tb]
  \includegraphics[width=8.5cm]{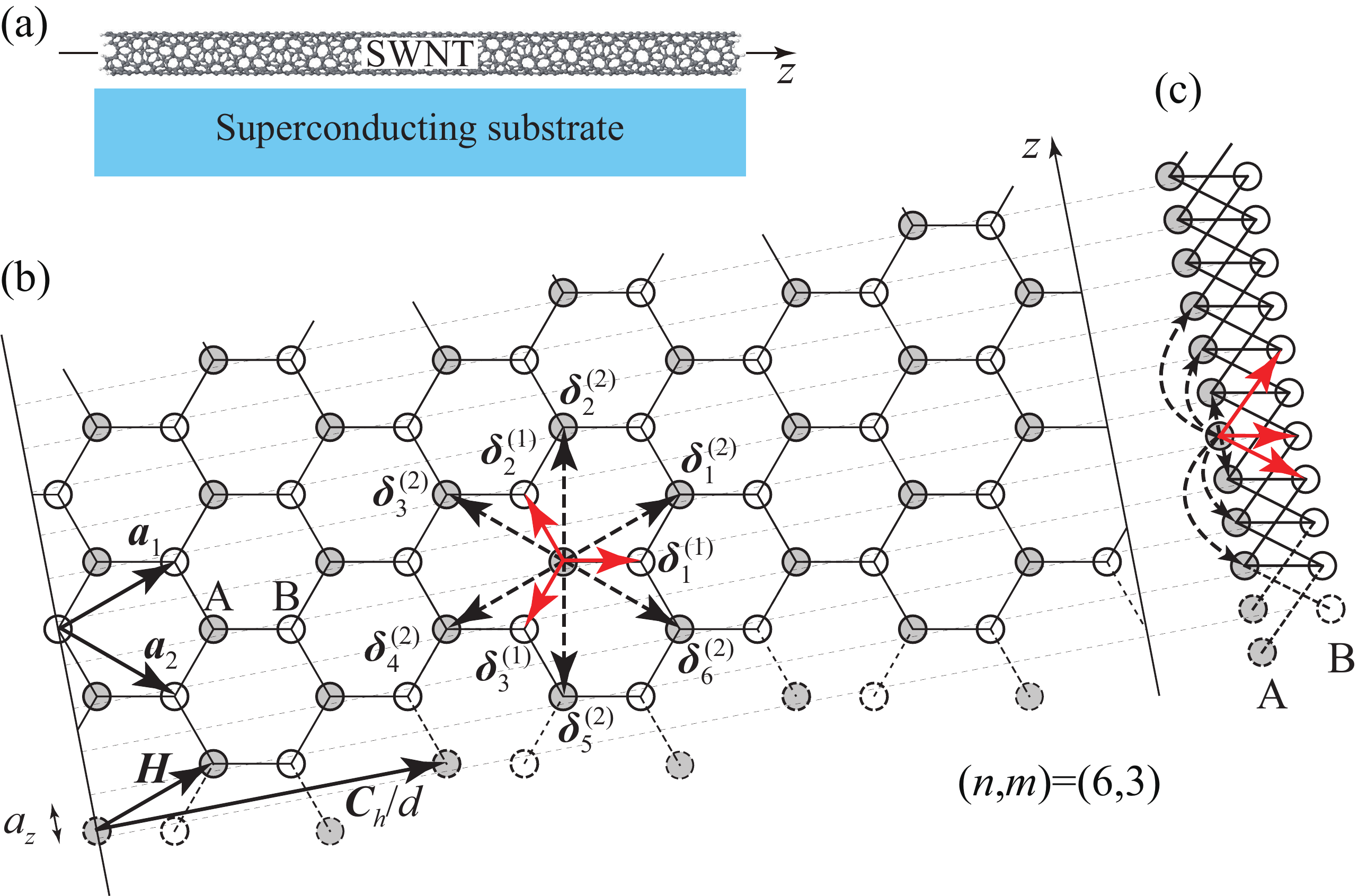}
  \caption{
    (a) Schematic figure of a SWNT proximity coupled to a
    superconducting substrate.
    (b) Hexagonal lattice structure.  Depicted are unit vectors $\mathbold{a}_1$,
    $\mathbold{a}_2$, alternative unit vectors $\mathbold{C}_h/d$,
    $\mathbold{H}$, and vectors to the nearest-neighbor and
    next-nearest-neighbor sites $\mathbold{\delta}_j^{(t)}$ for an
    unrolled $(n,m)=(6,3)$ SWNT, where $d=\mathrm{gcd}(n,m)=3$.
    A and B sublattices are denoted by gray and white circles,
    respectively.
    (c) An effective 1D lattice model, which is obtained by a partial
    Fourier transform in the circumferential direction, and is a
    projection of the 2D lattice structure onto the 1D
    nanotube axis $z$ (see the dashed lines).
    Solid lines denote nearest-neighbor bond connections in the
    original lattice structure.
}
  \label{fig:0603_2D1D}
\end{figure}

Proximitized SWNTs in appropriately tuned magnetic or electric fields
and with controlled gate voltage have been proposed as potential hosts
of edge states of Majorana nature.
Their formation relies on the spin-orbit coupling (either
native~\cite{Sau-PRB-2013}, or induced by 
an electric field~\cite{Klinovaja-PRL-2012},
a spiral magnetic
field~\cite{Egger-PRB-2012}, or a nuclear spin
helix~\cite{Hsu-PRB-2015-12}), as well as on breaking the
time-reversal symmetry.
In our model, we do not include an external magnetic field, thus the
time-reversal symmetry is preserved and the edge states always appear
in pairs.
In agreement with a recent work~\cite{Haim-PRB-2016}, we find no edge
states if only an onsite pairing is present.
Also, no edge states appear as long as the onsite pairing is larger
than
the nearest-neighbor one.
The inclusion of large nearest-neighbor pairings results in the
appearance of edge states which, interestingly, are just Dirac
fermions.

This paper is organized as follows.
In Sec.~\ref{sec:BdG_H_SWNT} a formulation for superconducting SWNTs
is given and the spectrum of the bulk system is presented.
In Sec.~\ref{sec:H_BdG_finite} the numerically calculated edge states
in the superconducting gap are shown and discussed.
In Sec.~\ref{sec:Winding} the winding number is introduced as a 1D
topological invariant and the topological phase diagram showing
regions of nontrivial winding number is given.
In Sec.~\ref{sec:Bulk-edge} a 1D continuum model is analyzed to show
the physics of the emergence of edge states and the bulk-edge
correspondence is proven.
In Sec.~\ref{sec:armchairClass} a case of strong valley coupling is
studied on the example of the armchair class SWNTs.  The conclusion is
given in Sec.~\ref{sec:Conclusion}.

\section{Bogoliubov--de Gennes Hamiltonian for finite-length SWNTs}
\label{sec:BdG_H_SWNT}

\subsection{Hamiltonian of a proximity-coupled SWNT}
\label{subsec:H_proximitySWNT}

Let us consider a SWNT proximity coupled to a superconducting
substrate [see Fig.~\ref{fig:0603_2D1D}(a)].
Proximity Hamiltonians have been investigated both in
graphene~\cite{Uchoa-PRL-2007,Burset-PRB-2008} and in
SWNTs~\cite{LeHur-PRB-2008}.
Following Ref.~\cite{Uchoa-PRL-2007}, we model the
$\pi$ electrons in the SWNT in terms of a tight-binding Hamiltonian,
which is given as the sum of a term $H_0$ describing the isolated
system and of a term $H_\mathrm{sc}$ accounting for proximity effects,
$H = H_0 + H_\mathrm{sc}$.
For later purpose, we first discuss some key features of the term
$H_0$ before turning to $H_\mathrm{sc}$.

A SWNT is defined by rolling up a graphene sheet in the direction of
the chiral vector $\mathbold{C}_h = n \mathbold{a}_1 + m
\mathbold{a}_2$, where $\mathbold{a}_1 = (\sqrt{3}/2, 1/2)a$ and
$\mathbold{a}_2 = (\sqrt{3}/2, -1/2)a$ are the unit vectors of
graphene, $a=0.246$~nm is the lattice constant, and the set of the two
integers $(n,m)$ defines the geometrical structure, called chirality,
of the SWNT~\cite{Saito-1998} [see Fig.~\ref{fig:0603_2D1D}(b)].
The term $H_0$, which includes curvature-induced
effects~\cite{Okuyama-2017-01}, is explicitly given as follows:
\begin{align}
  H_0
  = 
  & - \mu_\mathrm{c}
  \sum_{\mathbold{r} \sigma s}
  c_{\sigma \mathbold{r} s}^\dagger c_{\sigma \mathbold{r} s}  \nonumber \\
  & 
  + \sum_{\mathbold{r} s} 
  \sum_{j=1}^3 \gamma_{s,j}^{(1)}
  c_{\mathrm{A} \mathbold{r} s}^\dagger c_{\mathrm{B} \mathbold{r} + \mathbold{\delta}_j^{(1)} s} 
  + \mathrm{H.c.} 
  \nonumber \\
  & + \sum_{\mathbold{r} \sigma s} 
  \sum_{j=1}^3 \gamma_{s,j}^{(2)}
  c_{\sigma \mathbold{r} s}^\dagger c_{\sigma \mathbold{r} + \mathbold{\delta}_j^{(2)} s}
  + \mathrm{H.c.},
  \label{eq:H_0}
\end{align}
where $c_{\sigma \mathbold{r} s}$ is the annihilation operator of one
electron on sublattice $\sigma$ ($ = \mathrm{A, B}$) at site
$\mathbold{r}$ and with spin $s = \pm 1$.
The spin quantization axis is chosen to be the nanotube axis.
$\mu_\mathrm{c}$ sets the SWNT chemical potential and can be tuned,
possibly, through external gate voltages.
The vectors 
$\mathbold{\delta}_j^{(1)}$ ($j=1,2,3$) point to the three nearest-neighbor B sites from the A site,
and the vectors
$\mathbold{\delta}_j^{(2)}$ ($j=1,\cdots,6$) point to the six next-nearest-neighbor sites
[see Fig.~\ref{fig:0603_2D1D}(b)].
A spin-independent shift of the Dirac points is included in the nearest-neighbor hopping, 
while spin-orbit effects influence both the
nearest-neighbor and next-nearest-neighbor hoppings.
Reflecting the time-reversal symmetry we have $( \gamma_{-s,j}^{(t)} )^* = \gamma_{s,j}^{(t)}$.
The explicit forms of the vectors $\mathbold{\delta}_j^{(t)}$ and the
hopping integrals $\gamma_{s,j}^{(t)}$ ($t=1,2$) are provided in
Appendix~\ref{subsec:App:TBH_normal}.

Regarding the effective pairing Hamiltonian $H_\mathrm{sc}$, we notice
that the diameter $d_t$ of a SWNT is much smaller than a typical
superconducting penetration length $\lambda >
10$~nm~\cite{Tinkham-2004}.
Then, we can assume singlet superconducting pairing terms $\Delta_0$,
$\Delta_1$ being constant on the whole lattice, yielding~\cite{Uchoa-PRL-2007}
\begin{align}
  H_\mathrm{sc}
  & =
  \Delta_0 \left( \sum_{\mathbold{r} \sigma} 
  c_{\sigma \mathbold{r} \uparrow}^\dagger c_{\sigma \mathbold{r} \downarrow}^\dagger 
  + \mathrm{H.c.} \right) \nonumber \\
  & + 
  \Delta_1 \sum_{\mathbold{r}} \sum_{j=1}^3 
  \left( c_{\mathrm{A} \mathbold{r} \uparrow}^\dagger c_{\mathrm{B} \mathbold{r}+\mathbold{\delta}_j^{(1)} \downarrow}^\dagger 
  - c_{\mathrm{A} \mathbold{r} \downarrow}^\dagger c_{\mathrm{B} \mathbold{r}+\mathbold{\delta}_j^{(1)} \uparrow}^\dagger 
  + \mathrm{H.c.} \right).
\end{align}
Here, we have alternatively used $s =$ $\uparrow, \downarrow$ for the
spin index.
The term proportional to $\Delta_0$ represents the onsite pairing, and
the term proportional to $\Delta_1$ the pairing between the nearest-neighbor sites.
The gauge freedom allows us to choose the coupling terms $\Delta_0$, $\Delta_1$ as real numbers.
To determine the precise values of $\Delta_0$ and $\Delta_1$ for a
given chirality of SWNT contacted to a superconducting substrate, a
microscopic analysis of the interactions between the superconducting
substrate and the SWNT would be needed~\cite{LeHur-PRB-2008}, in
principle including also pairing correlations between next-nearest and
further neighbors.
However, as will be shown below, the presence of nearest-neighbor
pairing is the minimum requirement for the presence of nontrivial
topological phases.
Therefore in this paper we treat both $\Delta_0$ and $\Delta_1$ as
parameters in order to study their interplay.

\vspace*{-2mm}
\subsection{The 1D lattice Hamiltonian in the helical-angular construction}
\label{subsec:H_helicalAngular}

\begin{table}[tb]
\vspace*{-2mm}
  \caption{
    Hopping distance $\delta \ell_j^{(t)}$ and phase factor $\delta
    \nu_j^{(t)}$ in the 1D lattice
    model~\cite{Izumida-2016-05,Okuyama-2017-01}.
    The integers $p_s$ and $q_s$ satisfy $m p_s - n q_s = d$, where $d=\mathrm{gcd}(n,m)$.
  }
  \label{table:Deltaell}
  \begin{tabular}{ccccccc}
    \hline
    \hline
                         & $j=1$ & $j=2$ & $j=3$ & ~$j=4$ & $j=5$ & $j=6$ \\
    \hline
    $\delta \ell_j^{(1)}$ & $-\frac{n-m}{3d} $ & $\frac{2n+m}{3d}$ & $-\frac{2m+n}{3d}$ \\
    $\delta \nu_j^{(1)}$  & $\frac{p_s-q_s}{3}$ & $-\frac{2p_s+q_s}{3}$  & $\frac{2q_s+p_s}{3}$    \\
    $\delta \ell_j^{(2)}$ & $\frac{m}{d}$ & $\frac{n+m}{d}$ & $\frac{n}{d}$ & $-\frac{m}{d}$ & $-\frac{n+m}{d}$ & $-\frac{n}{d}$ \\
    $\delta \nu_j^{(2)}$  & $-q_s$ & $-(p_s+q_s)$ & $-p_s$ & $q_s$ & $p_s+q_s$ & $p_s$ \\
    \hline
    \hline
  \end{tabular}
\end{table}

Due to the $C_d$ rotational symmetry of a SWNT with respect to the
tube axis, the orbital angular momentum
$L_z = \hbar \mu$
is a well-defined quantity,
which is characterized by the integer,
\begin{equation}
  \mu = 0, 1, \cdots, d-1.
\end{equation}
Here, $d=\mathrm{gcd}(n,m)$ is the greatest common divisor of $n$ and
$m$.
Note that the angular momenta $\mu$ and $\mu'$ are equivalent if
$\mathrm{mod} (\mu - \mu',d) = 0$, thus, e.g., $\mu=-1$ is equivalent
to $\mu=d-1$.
Furthermore, also the spin component along the SWNT axis is a
conserved quantity, which allows us to decompose the Hamiltonian into
$\mathbold{\mu} \equiv (\mu,s)$ subspaces.
The decomposition is performed by a partial Fourier transform in the
circumference direction.
To achieve this, it is convenient to use the helical-angular
construction~\cite{White-1993-03,Izumida-2016-05,Okuyama-2017-01}, in
which the atomic position $\mathbold{r}$ is expressed by the
alternative unit vectors $\mathbold{C}_h / d$ and $\mathbold{H}$,
where $\mathbold{H}= p_s \mathbold{a}_1 + q_s \mathbold{a}_2$ with the
integers $p_s$ and $q_s$ satisfying $m p_s - n q_s = d$.
It holds 
$\mathbold{r} = \nu \left( \mathbold{C}_h / d \right) + \ell
\mathbold{H} + \delta_{\sigma,\mathrm{B}} \mathbold{\delta}_1^{(1)}$
with the two integers $\nu=0,1,\cdots,d-1$ and $\ell$.
The integer $\ell$ indicates the lattice position in the axis
direction in units of $a_z = \sqrt{3} a d / 2 \sqrt{n^2 + m^2 + nm}$,
which is the shortest distance between $\sigma$ atoms in the axis
direction [see Fig.~\ref{fig:0603_2D1D}(b)].
In this framework, the two-dimensional (2D) wave vector is expressed as 
$\mathbold{k} = \mu \mathbold{Q}_1 / d + k \mathbold{Q}_2 / (2 \pi/a_z)$, 
where $k$ is the wave number along the nanotube axis defined in the 1D Brillouin zone (BZ)
$-\pi/a_z \le k < \pi/a_z$, and $\mathbold{Q}_1$ and $\mathbold{Q}_2$
are the two reciprocal lattice vectors conjugated to $\mathbold{C}_h /
d$ and $\mathbold{H}$, respectively.
That is, the relations
$\mathbold{Q}_1 \cdot \mathbold{C}_h / d = \mathbold{Q}_2 \cdot \mathbold{H} = 2 \pi$ and 
$\mathbold{Q}_1 \cdot \mathbold{H} = \mathbold{Q}_2 \cdot \mathbold{C}_h /d = 0$ hold.
Then, the partial Fourier transform is expressed as
\begin{equation}
  c_{\sigma \mathbold{r} s} 
  = 
  \frac{1}{\sqrt{d}} \sum_{\mu=0}^{d-1} 
  \exp \left( i \frac{2 \pi}{d} \nu \mu \right)
  c_{\sigma \ell \mathbold{\mu}}.
  \label{eq:partialFourier}
\end{equation}
The Hamiltonian of the normal term is rewritten as $H_0 =
\sum_{\mathbold{\mu}} H_{0,\mathbold{\mu}}$,
where~\cite{Izumida-2015-06,Izumida-2016-05,Okuyama-2017-01},
\begin{align}
  H_{0, \mathbold{\mu}} 
  = 
  & -\mu_\mathrm{c} \sum_{\ell \sigma} 
  c_{\sigma \ell \mathbold{\mu}}^\dagger c_{\sigma \ell \mathbold{\mu}} \nonumber \\
  & + 
  \sum_{\ell} \sum_{j=1}^3 
  e^{i \frac{2 \pi}{d} \delta \nu_j^{(1)} \mu}
  \gamma_{s,j}^{(1)}
  c_{\mathrm{A} \ell \mathbold{\mu}}^\dagger
  c_{\mathrm{B} \ell_j^{(1)} \mathbold{\mu}} 
  + \mathrm{H.c.} \nonumber \\
  & +
  \sum_{\ell \sigma} \sum_{j=1}^3 
  e^{i \frac{2 \pi}{d} \delta \nu_j^{(2)} \mu}
  \gamma_{s,j}^{(2)}
  c_{\sigma \ell \mathbold{\mu}}^\dagger
  c_{\sigma \ell_j^{(2)} \mathbold{\mu}}
  + \mathrm{H.c.},
 \label{eq:H_0_1Dlattice}
\end{align}
where
\begin{equation}
  \ell_j^{(t)} = \ell + \delta \ell_j^{(t)}, \quad t=1,2.
\end{equation}
The hopping distance $\delta \ell_j^{(t)}$ and the phase factor
$\delta \nu_j^{(t)}$ are determined from $\mathbold{\delta}_j^{(t)} =
\delta \nu_j^{(t)} \mathbold{C}_h / d + \delta \ell_j^{(t)}
\mathbold{H}$.
Their explicit expressions are given in Table~\ref{table:Deltaell}.
As schematically shown in Fig.~\ref{fig:0603_2D1D}(c), the Hamiltonian
in each $\mathbold{\mu}$ subspace represents a ladder-type 1D lattice
model~\cite{White-1993-03,Izumida-2016-05,Okuyama-2017-01}.

Under the partial Fourier transform of Eq.~\eqref{eq:partialFourier},
the superconducting term of the Hamiltonian takes the form,
\begin{align}
  H_\mathrm{sc}
& = 
  \sum_{\mathbold{\mu}} 
  \Biggl[
    \frac{\Delta_0}{2}
    \sum_{\ell \sigma} 
    s c_{\sigma \ell \mathbold{\mu}}^\dagger
  c_{\sigma \ell -\mathbold{\mu}}^\dagger
  + \mathrm{H.c.} \nonumber \\
&
  + 
  \Delta_1 
  \sum_{\ell} \sum_{j=1}^3 
  e^{i \frac{2 \pi}{d} \delta \nu_j^{(1)} \mu}
  s c_{\mathrm{A} \ell \mathbold{\mu}}^\dagger
  c_{\mathrm{B} \ell_j^{(1)} -\mathbold{\mu}}^\dagger
  + \mathrm{H.c.}
  \Biggr].
 \label{eq:H_s_1Dlattice}
\end{align}
The pair $\mathbold{\mu}$ and $-\mathbold{\mu}$ in the superconducting
term reflects the conservation of angular momentum and spin.

\subsection{Bogoliubov--de Gennes formalism for the 1D lattice Hamiltonian}
\label{subsec:H_BdG_helicalAngular}

Since the total Hamiltonian $H_0 + H_\mathrm{sc}$ has a bilinear form
in the fermionic operators $c_{\sigma \ell \mathbold{\mu}}$, the
excitation spectrum is conveniently calculated within the
Bogoliubov--de Gennes (BdG) formalism~\cite{Tinkham-2004}.
The BdG Hamiltonian $\mathcal{H}$ is given by doubling the fermionic
operators upon introduction of the Nambu spinor
\begin{equation}
  \mathbf{c}_{\sigma \ell \mathbold{\mu}}^\dagger
  =
  \left(
    c_{\sigma \ell \mathbold{\mu}}^\dagger,
    c_{\sigma \ell -\mathbold{\mu}}
  \right),
  \quad
  \mathbf{c}_{\sigma \ell \mathbold{\mu}}
  =
  \left(
  \begin{array}{c}
    c_{\sigma \ell \mathbold{\mu}} \\
    c_{\sigma \ell -\mathbold{\mu}}^\dagger
  \end{array}
  \right).
  \label{eq:NambuSpinor}
\end{equation}
For instance, the superconducting term proportional to $\Delta_0$ in
Eq.~\eqref{eq:H_s_1Dlattice} is rewritten as
\begin{equation}
  \Delta_0 \sum_{s} s 
  c_{\sigma \ell \mathbold{\mu}}^\dagger
  c_{\sigma \ell -\mathbold{\mu}}^\dagger + \mathrm{H.c.}
 =
  \Delta_0 \sum_{s} s 
  \mathbf{c}_{\sigma \ell \mathbold{\mu}}^\dagger
  \hat{\pi}_x
  \mathbf{c}_{\sigma \ell \mathbold{\mu}}.
\end{equation}
Here we have introduced the Pauli matrices 
$(\hat{\pi}_x, \hat{\pi}_y, \hat{\pi}_z)$ acting in the particle-hole
subspace.
Detailed transformation to the BdG form of the superconducting term
proportional to $\Delta_1$ in Eq.~\eqref{eq:H_s_1Dlattice} is given in
Appendix~\ref{subsec:App:BdG_Delta1}.
Collecting all terms, the BdG Hamiltonian for the SWNTs is expressed as $\mathcal{H} =
\frac{1}{2} \sum_{\mathbold{\mu}} \mathcal{H}_{\mathbold{\mu}}$, where
\begin{align}
  \mathcal{H}_{\mathbold{\mu}} 
  &
  = 
  \sum_{\ell \sigma} 
  \mathbf{c}_{\sigma \ell \mathbold{\mu}}^\dagger
  \left( -\mu_\mathrm{c} \hat{\pi}_z + s \Delta_0 \hat{\pi}_x \right)
  \mathbf{c}_{\sigma \ell \mathbold{\mu}} \nonumber \\
  & 
  + \left[
    \sum_{\ell} \sum_{j=1}^3 
    e^{i \frac{2 \pi}{d} \delta \nu_j^{(1)} \mu}
    \mathbf{c}_{\mathrm{A} \ell \mathbold{\mu}}^\dagger
    \left( \gamma_{s,j}^{(1)} \hat{\pi}_z + s \Delta_1 \hat{\pi}_x \right)
    \mathbf{c}_{\mathrm{B} \ell_j^{(1)} \mathbold{\mu}} 
    \right. \nonumber \\
  & \left. 
    +
    \sum_{\ell \sigma} \sum_{j=1}^3 
    e^{i \frac{2 \pi}{d} \delta \nu_j^{(2)} \mu}
    \gamma_{s,j}^{(2)} 
    \mathbf{c}_{\sigma \ell \mathbold{\mu}}^\dagger
    \hat{\pi}_z
    \mathbf{c}_{\sigma \ell_j^{(2)} \mathbold{\mu}}
    + \mathrm{H.c.} \right].
  \label{eq:H_BdG_1Dlattice}
\end{align}
In each $\mathbold{\mu}$ subspace $\mathcal{H}_\mathbold{\mu}$ represents a 1D
ladder Hamiltonian, which extends to the BdG form a previously developed 1D
lattice model for the normal
state~\cite{Izumida-2015-06,Izumida-2016-05,Okuyama-2017-01}.

The doubling also gives a particle-hole symmetry to the BdG excitation spectrum.
The BdG spectrum in a finite-length SWNT with $\ell =
1,2,\cdots,N_L$ lattice sites is numerically calculated by
diagonalizing the Hamiltonian in Eq.~\eqref{eq:H_BdG_1Dlattice}, and
will be analyzed in Sec.~\ref{sec:H_BdG_finite}.
Before doing this, we discuss the BdG spectrum of the bulk system.

\subsection{Energy bands and BdG spectrum of the bulk system}
\label{subsec:H_En_BdG_bulk}

Exploiting translational invariance, the BdG Hamiltonian of the bulk
system is written in the Bloch basis as
$\mathcal{H}_{\mathbold{\mu}} 
= \sum_{k} \mathbf{c}_{k \mathbold{\mu}}^\dagger \mathcal{H}_{\mathbold{\mu}}(k) \mathbf{c}_{k \mathbold{\mu}}$,
where
\begin{align}
  \mathcal{H}_{\mathbold{\mu}}(k) 
  = & 
    \left(
    \begin{array}{cc}
      \varepsilon_{\mathrm{c},\mathbold{\mu}}(k) & f_{\mathrm{e},\mathbold{\mu}}(k) \\
      f_{\mathrm{e},\mathbold{\mu}}^*(k)   & \varepsilon_{\mathrm{c},\mathbold{\mu}}(k)
    \end{array}
    \right) \hat{\pi}_z \nonumber \\
    & +
    s \left(
    \begin{array}{cc}
      \Delta_0               & f_{\mathrm{eh},\mu}(k) \\
      f_{\mathrm{eh},\mu}^*(k) & \Delta_0
    \end{array}
    \right) \hat{\pi}_x,
    \label{eq:H_BdG_bulk_0}
\end{align}
and
\begin{align}
  & f_{\mathrm{e},\mathbold{\mu}}(k) 
  = \sum_{j=1}^3 \gamma_{s,j}^{(1)} e^{i \mathbold{k} \cdot \mathbold{\delta}_j^{(1)}}, \quad
  f_{\mathrm{eh},\mu}(k) = \Delta_1 \sum_{j=1}^3 e^{i \mathbold{k} \cdot \mathbold{\delta}_j^{(1)}}, 
  \nonumber \\
  & 
  \varepsilon_{\mathrm{c},\mathbold{\mu}}(k)
  = - \mu_{\mathrm{c}} + \varepsilon_{\mathrm{so},\mathbold{\mu}}(k), 
\quad
  \varepsilon_{\mathrm{so},\mathbold{\mu}}(k) 
  = \sum_{j=1}^6 \gamma_{s,j}^{(2)} e^{i \mathbold{k} \cdot \mathbold{\delta}_j^{(2)}}.
\end{align}
The Nambu spinor in $k$ space is
$\mathbf{c}_{k \mathbold{\mu}}^\dagger =
\left(
c_{\mathrm{A} k \mathbold{\mu}}^\dagger,
c_{\mathrm{B} k \mathbold{\mu}}^\dagger,
c_{\mathrm{A} -k -\mathbold{\mu}},
c_{\mathrm{B} -k -\mathbold{\mu}} \right)$
with 
$c_{\sigma k \mathbold{\mu}} 
= 
\frac{1}{\sqrt{N_L}} \sum_{\ell}
\exp \left( - i k a_z \ell \right)
c_{\sigma \ell \mathbold{\mu}}$,
and
$\mathbold{k} = (\mu,k)$.
The BdG spectrum of the bulk system is obtained by diagonalizing the
Hamiltonian matrix of Eq.~\eqref{eq:H_BdG_bulk_0}.

\subsubsection{Energy bands for the normal case}

Before showing the BdG spectrum of the bulk system, we shall review the
energy bands of the normal case.
Until discussing the BdG spectrum, we set the chemical potential to be
zero, $\mu_\mathrm{c} = 0$.
The conduction and the valence bands of the SWNTs are given by
diagonalizing the matrix in the first term of
Eq.~\eqref{eq:H_BdG_bulk_0} and have the standard form~\cite{Laird-RevModPhys-2015}
\begin{equation}
  \varepsilon_{\mathbold{\mu}}(k) =
  \varepsilon_{\mathrm{so},\mathbold{\mu}}(k) \pm
  |f_\mathrm{e,\mathbold{\mu}}(k)|,
  \label{eq:eband}
\end{equation}
where the signs $+$ and $-$ correspond to the conduction and the
valence bands, respectively.

It is well known that the SWNTs are metallic when $\mathrm{mod} (2n +
m, 3) = 0$ and semiconducting if $\mathrm{mod} (2n + m, 3) =
1,2$~\cite{Saito-1998}.
Recent
studies~\cite{Lunde-PRB-2005,Izumida-2015-06,Marganska-PRB-2015,Izumida-2016-05}
have revealed that the SWNTs can be alternatively classified into two
classes according to the angular momentum of the two valleys, denoted
in the following $K$ and $K'$:
(i) zigzag class, which includes metal-1 (metallic SWNTs with $d_R =
d$) and semiconducting SWNTs with $d \ge 4$, in which the two valleys
have different angular momenta, where $d_R=\mathrm{gcd}(2n+m,2m+n)$;
(ii) armchair class, which includes metal-2 (metallic SWNTs with $d_R
= 3d$) and semiconducting SWNTs with $d \le 2$, in which the two
valleys have the same angular momentum.
Here the angular momentum $\mu_\tau$ of valley $\tau$ $(= K, K')$ is
defined as follows.
For the metallic SWNTs, $\mu_\tau$ is the angular momentum at the
$\tau$ point, which is given by $\mu_\tau = \mathrm{mod} \left[ \tau
  (2n+m)/3, d \right]$~\cite{Izumida-2016-05}, where we have
alternatively used $\tau = 1$ ($-1$) for the valley $K$ ($K'$).
At the same time, the 1D wave number for the $\tau$ point is given by
$k_\tau = (2 \pi/3 a_z) \mathrm{mod} \left[ \tau (2p + q), 3
  \right]$~\cite{Izumida-2016-05}.
For the semiconducting SWNTs, the corresponding angular momenta and
the 1D wave numbers are given by the ones which are closest to the
$\tau$ point.
Their explicit expressions are also given in
Ref.~\cite{Izumida-2016-05}.

\begin{figure}[tb]
  \includegraphics[width=8.5cm]{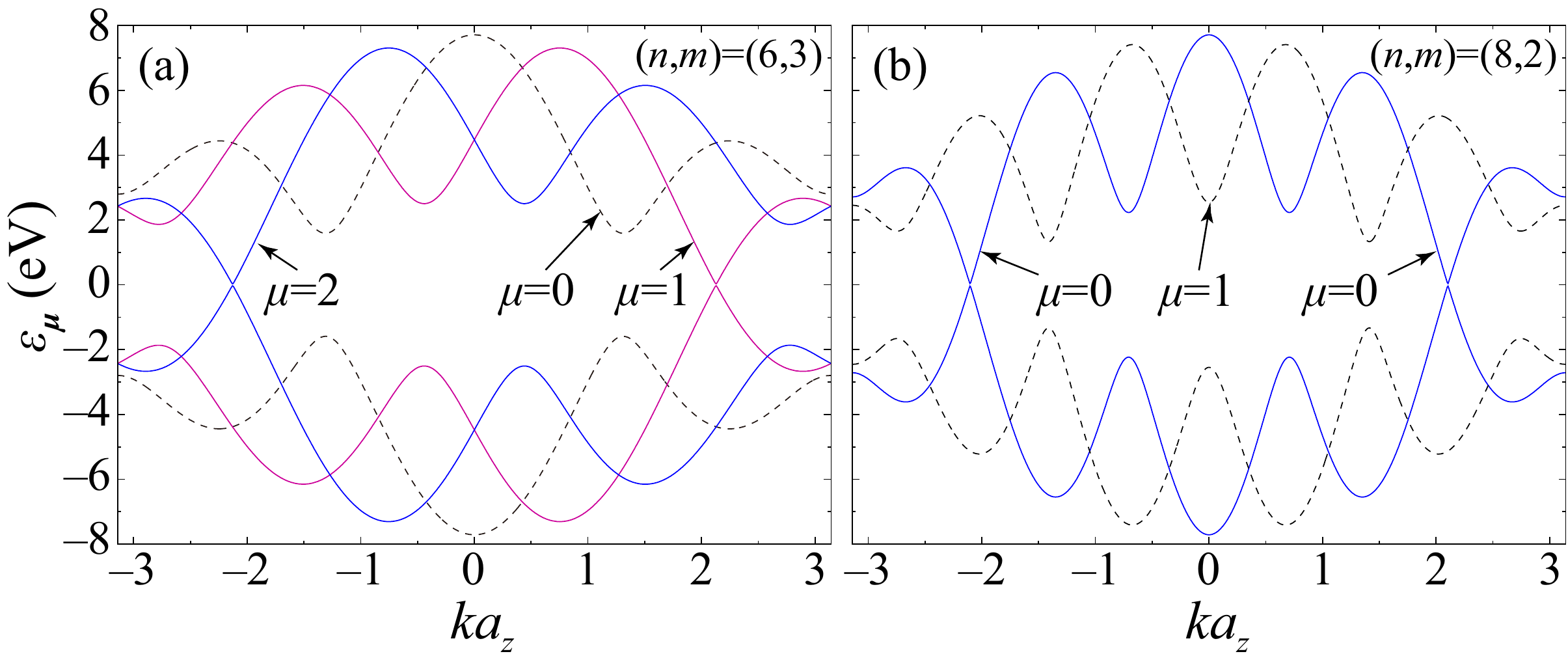}
  \caption{
    Conduction and valence bands of (a) $(n,m)=(6,3)$ metal-1 SWNT,
    classified into the zigzag class, and (b) $(n,m)=(8,2)$ metal-2
    SWNT, classified into the armchair class.
    For both cases (a) and (b), $p_s=1$ and $q_s=0$.
    The angular momentum for each band is indicated, the blue curves
    show bands with $\mu=\mu_K$, and the purple curves in (a) show bands with $\mu=\mu_{K'}$.
    Curvature-induced energy gaps at zero energy and spin-orbit
    splitting are not seen on this energy scale.
  }
  \label{fig:Enk}
\end{figure}

Specifically, $\mu_K=\mu_{K'}=0$ holds for the metal-2
SWNTs~\cite{Izumida-2015-06}.
Figure~\ref{fig:Enk} clearly shows the above features:
in Fig.~\ref{fig:Enk}(a) we depict the energy bands of an $(n,m)=(6,3)$ SWNT
which belongs to the zigzag class.
The angular momentum $\mu_K=2$ of valley $K$ is different from that of
the $K'$ valley which is $\mu_{K'}=1$.
On the other hand, Fig.~\ref{fig:Enk}(b) shows the energy band of an
$(n,m)=(8,2)$ SWNT, representative of the armchair class, where
$\mu_K=\mu_{K'}=0$.

Since we are focusing on the states near the $K$ and $K'$ valleys, it
is sufficient to consider a limited number of $\mathbold{\mu}$
subspaces, which are specified by the angular momenta of the valleys.
In the following, we focus on the zigzag class SWNTs where two
valleys are well decoupled.
The armchair class will be discussed in Sec.~\ref{sec:armchairClass}.

\subsubsection{BdG spectrum of zigzag class SWNTs}

\begin{figure}[tb]
  \includegraphics[width=8.5cm]{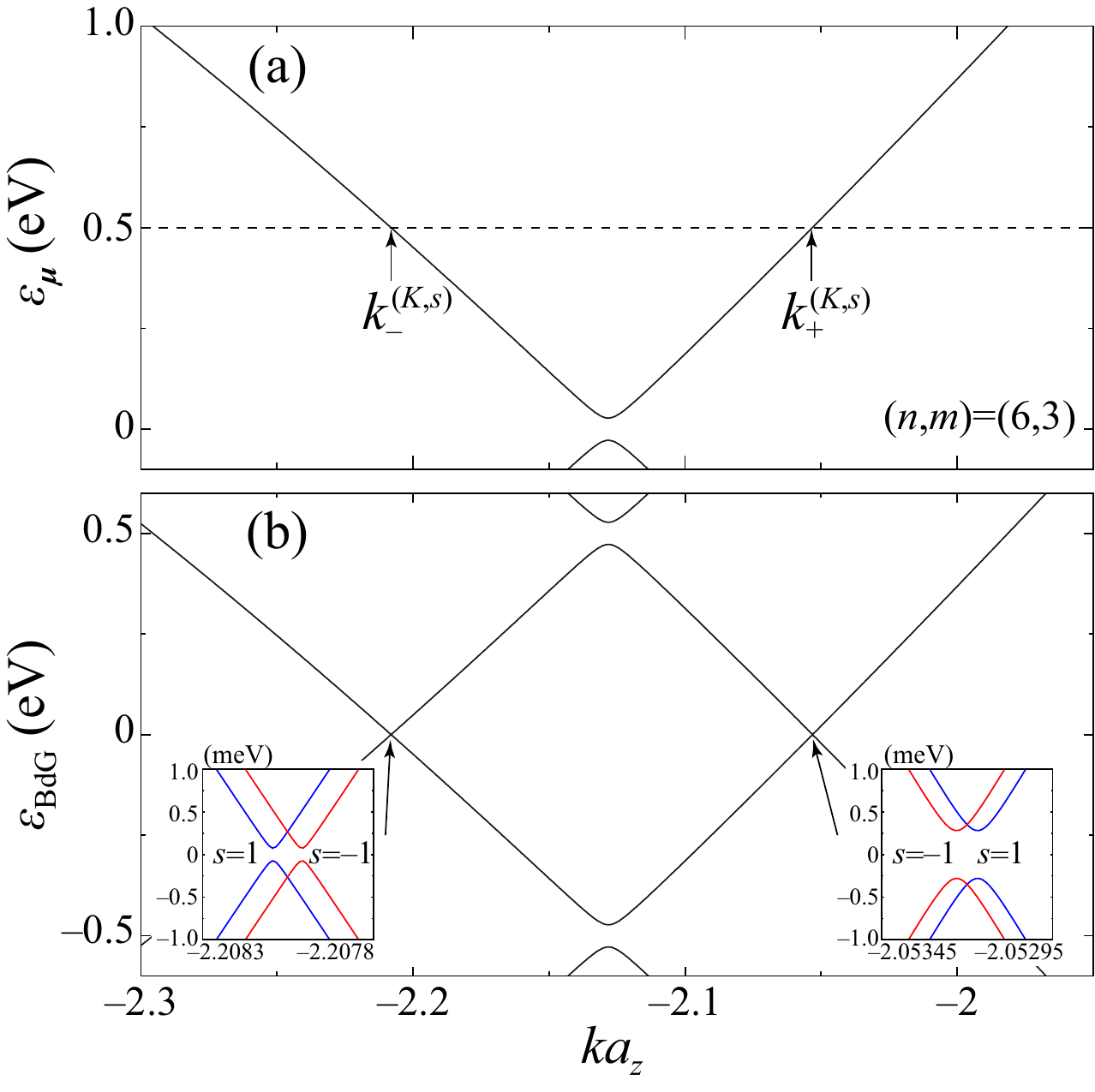}
  \caption{
    (a) Energy band, and (b) BdG excitation spectrum near the $K$
    point for an $(n,m)=(6,3)$ SWNT.
    The chemical potential is set to be $\mu_\mathrm{c} = 500$~meV.
    The two arrows with $k_\pm^{(K,s)}$ in (a) indicate the two Fermi points.
    In (b) the superconducting coupling parameters are chosen to be
    $\Delta_0 = 0.5$~meV and $\Delta_1 = 2$~meV.
    Each inset in (b) shows the enlarged BdG spectrum near the two Fermi
    points.
    The blue and red curves show the BdG spectra for the spin-up and~-down
    components, respectively.
}
  \label{fig:0603_Esck}
\end{figure}

Since our interest is on the impact of superconductivity on the
conducting electrons, the chemical potential will be set in the energy
region corresponding to electron transport.
Figure~\ref{fig:0603_Esck}(a) shows the energy bands near the $K$
point for an $(n,m)=(6,3)$ SWNT.
The dashed line indicates the chemical potential.
Figure~\ref{fig:0603_Esck}(b) shows the corresponding BdG spectrum.
As shown in the two insets, the BdG spectrum exhibits small
superconducting gaps of the order of the superconducting couplings
near the two Fermi points $k = k_-^{(\mathbold{\tau})}$ and
$k=k_+^{(\mathbold{\tau})} (> k_-^{(\mathbold{\tau})})$, at which
\begin{equation}
  \mu_\mathrm{c} = \varepsilon_{\mathbold{\mu}_\tau}(k_r^{(\mathbold{\tau})}), \quad (r=\pm 1),
\end{equation}
is satisfied in the $\tau$ valley, where $\mathbold{\tau} \equiv
(\tau,s)$, $\mathbold{\mu}_\tau \equiv (\mu_\tau,s)$, and 
$\varepsilon_{\mathbold{\mu}_\tau}(k_r^{(\mathbold{\tau})})$ is the
single-particle energy of band $\mathbold{\mu}_\tau$ at
$k_r^{(\mathbold{\tau})}$ given by Eq.~\eqref{eq:eband}.

For a moderate chemical potential $|\mu_\mathrm{c}| \lesssim 1$~eV, the
$\mathbold{k}\cdot\mathbold{p}$ scheme can be used.
The hopping functions $f_{\mathrm{e},\mathbold{\mu}}$ and
$f_{\mathrm{eh},\mathbold{\mu}}$ are expanded around the $\tau$ point
as~\cite{Izumida-2009-06}
\begin{align}
  & f_{\mathrm{e},\mathbold{\mu}}(k) \simeq 
  c \gamma \left[ \left( k_z - \tau \Delta k_z \right) + i \left( k_c - \Delta k_{c,\mathbold{\tau}} \right) \right], \nonumber \\
  & f_{\mathrm{eh},\mu}(k) \simeq c \Delta_1 \left( k_z + i k_c \right),
  \quad
  \varepsilon_{\mathrm{c},\mathbold{\mu}}(k) \simeq - \mu_{\mathrm{c},\mathbold{\tau}},
  \label{eq:f_expansion}
\end{align}
where $c$ is a complex coefficient 
encoding the chiral angle and valley
with $|c| = \sqrt{3}a/{2}$, and
\begin{equation}
  \Delta k_{c,\mathbold{\tau}} = \Delta k_{c} + \tau s \Delta k_{\mathrm{so}}, \quad
  \mu_{\mathrm{c},\mathbold{\tau}} = \mu_{\mathrm{c}} - \tau s \varepsilon_{\mathrm{so}}.
  \label{eq:Deltak_muc_SO}
\end{equation}
Here, $\Delta k_z$, $\Delta k_c$ are the curvature-induced shifts of the
Dirac point from the $K$ point in the circumferential and the axial
directions, respectively.
$\Delta k_\mathrm{so}$ and $\varepsilon_\mathrm{so}$ are the
spin-dependent Dirac point shift in the circumferential direction and
the Zeeman-type energy shift, respectively, induced by the spin-orbit
interaction.
Their explicit expressions are given in
Eqs.~\eqref{eq:app:Deltakc_Deltakz} and \eqref{eq:app:Deltakso_epsso} in
Appendix~\ref{subsec:App:TBH_normal}.
Finally, $k_z$ and $k_c$ are the wave numbers in the circumferential and
axial directions measured from the $\tau$ point, and $k_c = 0$,
$-2/3d_t$, and $2/3d_t$ for metallic, type-1 [$\mathrm{mod} (2n + m, 3)
  = 1$], and type-2 [$\mathrm{mod} (2n + m, 3) = 2$] semiconducting
SWNTs, respectively.
Using Eq.~\eqref{eq:f_expansion}, the two Fermi points measured from
the $\tau$ point are given by,
\begin{equation}
  k_{r}^{(\mathbold{\tau})} = \tau \Delta k_z 
  + r \sqrt{ \left( \frac{\mu_{\mathrm{c},\mathbold{\tau}}}{|c| \gamma} \right)^2 - \left( k_c - \Delta k_{c,\mathbold{\tau}} \right)^2 }.
  \label{eq:k_r}
\end{equation}
And, as shown in Appendix~\ref{sec:App:BdGspectrum}, the
superconducting gap near $k_{r}^{(\mathbold{\tau})}$ is expressed as
\begin{align}
  \varepsilon_{\mathrm{g}, r}^{(\mathbold{\tau})}
  = &
  2 \Delta_0 
  + 2 \Delta_1 \frac{\mu_{\mathrm{c}, \mathbold{\tau}}}{\gamma}
  \Bigl[ 1 + \varepsilon_{c, \mathbold{\tau}} E_{c, \mathbold{\tau}} \nonumber \\
&
  - r \tau \sgn (\mu_{\mathrm{c}, \mathbold{\tau}}) \varepsilon_{z,\mathbold{\tau}} \sqrt{1 - E_{c, \mathbold{\tau}}^2} 
  \Bigr],
  \label{eq:egap}
\end{align}
where
\begin{align}
  & E_{c,\mathbold{\tau}} = \frac{|c| \gamma \left( k_c - \Delta k_{c,\mathbold{\tau}} \right)}{\mu_{\mathrm{c},\mathbold{\tau}}}, \quad 
  \varepsilon_{c,\mathbold{\tau}} = \frac{|c| \gamma \Delta k_{c,\mathbold{\tau}}}{\mu_{\mathrm{c},\mathbold{\tau}}}, \nonumber \\
  & \varepsilon_{z,\mathbold{\tau}} = \frac{|c| \gamma \Delta k_z}{\mu_{\mathrm{c},\mathbold{\tau}}}.
  \label{eq:def_Ec}
\end{align}
Since the absolute value of the numerator of $E_{c,\mathbold{\tau}}$
expresses the half of the bulk band gap, the relation
$|E_{c,\mathbold{\tau}}| < 1$ holds when the chemical potential is in
the energy band regions.
It should be noted that the superconducting gaps at the two Fermi
points $k_{r}^{(\mathbold{\tau})}$ ($r= \pm$) are different as shown
in the inset of Fig.~\ref{fig:0603_Esck}(b) as well as expressed in
Eq.~\eqref{eq:egap}.
This is because the contribution of $\Delta_1$ to the superconducting
gap is $k$ dependent, as shown in Eq.~\eqref{eq:f_expansion}, and
the contribution at the two Fermi points is different, reflecting the
shift $\Delta k_z$ of the Dirac point.
The two different superconducting gaps at the two Fermi points play an
important role in the emergence of edge states, as will be discussed
later.

Next, we focus on the low-energy BdG excitations, of the order of the
superconducting gaps, in finite-length SWNTs.

\section{B\lowercase{d}G spectrum in finite-length SWNTs}
\label{sec:H_BdG_finite}

\begin{figure}[t]
  \includegraphics[width=8.5cm]{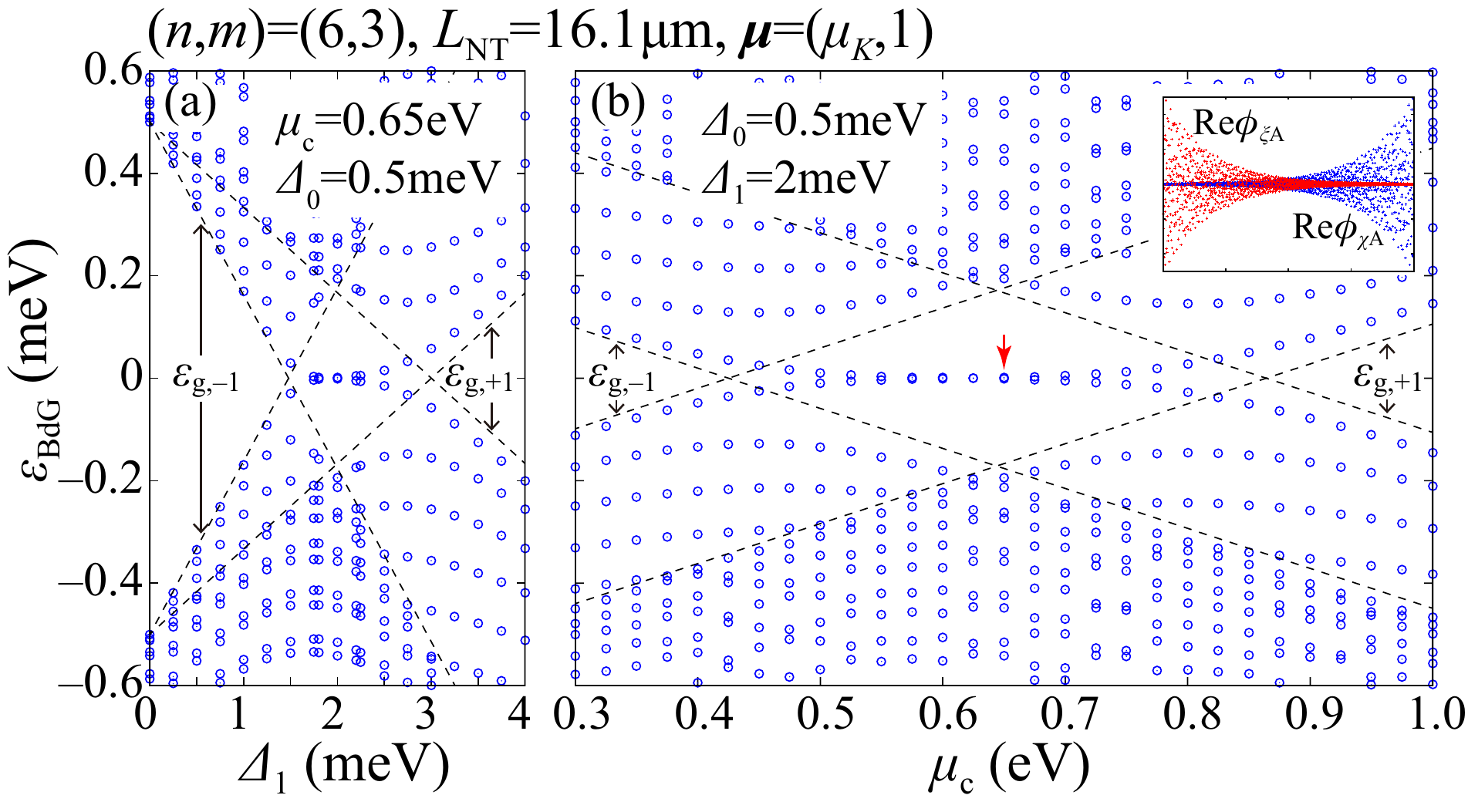}
  \caption{
    BdG spectrum of a $(6,3)$ nanotube with a length of $16.1$~$\mu$m
    in the $\mathbold{\mu}=(\mu_K=2,1)$ subspace.
    (a) Spectrum as a function of the superconducting pairing
    $\Delta_1$, and (b) as a function of the chemical potential
    $\mu_\mathrm{c}$.
    Blue circles show the calculated spectrum and the dashed lines
    show the superconducting gaps $\varepsilon_{\mathrm{g},
      r}^{(K,s=1)}$ of the bulk system given in Eq.~\eqref{eq:egap}.
    The inset in (b) shows the real part components $\phi_{\chi
      \mathrm{A}}$ (blue) and $\phi_{\xi \mathrm{A}}$ (red) 
    in arbitrary units as a function of lattice site $\ell$ for
    the
    calculated eigenfunction at $\varepsilon_{\mathrm{BdG}}=0$ with
    $\Delta_0 = 0.5$~meV, $\Delta_1 = 2$~meV and $\mu_\mathrm{c} =
    650$~meV [indicated by the red arrow in (b)].
    The definition of $\phi_{p \sigma}$ ($p = \chi, \xi$) is given in
    Eq.~\eqref{eq:phi_chi_xi}.
}
  \label{fig:0603_BdG_Finite}
\end{figure}

We focus on an $(n,m)=(6,3)$ SWNT with $N_L = 2 \times 10^5$, which
corresponds to a SWNT length of $16.1$~$\mu$m, as an example for the
zigzag class SWNTs.
The BdG Hamiltonian is diagonalized as
\begin{equation}
  \mathcal{H}_{\mathbold{\mu}} 
  = \sum_{l_v} \varepsilon_{\mathrm{BdG}}^{(\mathbold{\mu} l_v)}
  b_{\mathbold{\mu} l_v}^\dagger b_{\mathbold{\mu} l_v},
\end{equation}
where $l_v$ enumerates the quasiparticle energy levels, and
\begin{equation}
  b_{\mathbold{\mu} l_v}^\dagger 
  = 
  \sum_{\ell \sigma} 
  \left( \phi_{\mathrm{p} \sigma}^{(\mathbold{\mu} l_v)}(\ell) c_{\sigma \ell \mathbold{\mu}}^\dagger
  + \phi_{\mathrm{h} \sigma}^{(\mathbold{\mu} l_v)}(\ell) c_{\sigma \ell -\mathbold{\mu}}
  \right).
\end{equation}
Figure~\ref{fig:0603_BdG_Finite} shows the calculated spectrum in the
energy region of the order of the superconducting gaps in the subspace
$\mathbold{\mu}=(\mu_K=2,1)$.
The boundary shape is depicted in Fig. \ref{fig:0603_2D1D}(c), which
belongs to the class of so-called minimal boundaries.
The eigenvalue solver FEAST~\cite{Polizzi-FEAST-2009} of the Intel
Math Kernel Library was used for the numerical calculation.
The dashed lines show the evolution of the superconducting gaps
$\varepsilon_{\mathrm{g}, r}^{(\mathbold{\tau})}$ given in
Eq.~\eqref{eq:egap} with $\Delta_1$
[Fig.~\ref{fig:0603_BdG_Finite}(a)] and $\mu_\mathrm{c}$
[Fig.~\ref{fig:0603_BdG_Finite}(b)].
The functions $\phi_{\chi \mathrm{A}}$, $\phi_{\xi \mathrm{A}}$ shown
in the inset of Fig.~\ref{fig:0603_BdG_Finite}(b) are connected to
$\phi_{\mathrm{p} \mathrm{A}}$, $\phi_{\mathrm{h} \mathrm{A}}$ by a
unitary transformation
\begin{equation}
  \left(
  \begin{array}{c}
    \phi_{\chi \sigma} \\
    \phi_{\xi \sigma}
  \end{array}
  \right)
  = U_\pi^{-1}
  \left(
  \begin{array}{c}
    \phi_{\mathrm{p} \sigma} \\
    \phi_{\mathrm{h} \sigma}
  \end{array}
  \right),
  \label{eq:phi_chi_xi}
\end{equation}
where
\begin{equation}
  U_\pi = 
  \frac{1}{2} 
  \left(
  \begin{array}{cc}
     1 + i & 1 + i \\
    -1 + i & 1 - i
  \end{array}
  \right).
  \label{eq:U_pi}
\end{equation}
We will discuss this transformation in the next section.
In the region $1.5 \lesssim \Delta_1 \lesssim 3$~meV in Fig.~\ref{fig:0603_BdG_Finite}(a) 
and $400
\lesssim \mu_c \lesssim 900$~meV in Fig.~\ref{fig:0603_BdG_Finite}(b), 
states near zero energy exist
inside the gap region.
As shown in the inset in Fig.~\ref{fig:0603_BdG_Finite}(b), these
states are localized at the edges and their nature will be discussed
in the coming sections.
Calculations for the other three subspaces, $(\mu_K,-1)$ and
$(\mu_{K'},\pm1)$, where $\mu_{K'}=1$, exhibit an almost identical
behavior (not shown) as the one seen in
Fig.~\ref{fig:0603_BdG_Finite}.
Emergence of the zero energy states in these region is also seen (not
shown) for other boundary shapes, e.g., when removing or adding a A
sublattice at the end of the boundary shown in
Fig. \ref{fig:0603_2D1D}(c).

The numerical result in Fig.~\ref{fig:0603_BdG_Finite} clearly shows
that there exist edge states at zero energy in some parameter regions.
To explore the condition for the emergence of the edge states, we will
analyze the bulk system from a topological viewpoint.

\section{Winding number}
\label{sec:Winding}

Let us again consider the bulk Hamiltonian given in
Eq.~\eqref{eq:H_BdG_bulk_0}.
Since the Hamiltonian has the chiral symmetry $\{ \Gamma,
\mathcal{H}_{\mathbold{\mu}} \} = 0$, where $\Gamma = \hat{\pi}_y$,
one can introduce the winding number
\begin{equation}
  w_{\mathbold{\mu}} 
  = -\frac{1}{4 \pi i} 
  \int_{-\pi/a_z}^{\pi/a_z} dk 
  \mathrm{Tr} \left[ \Gamma \mathcal{H}^{-1}_{\mathbold{\mu}}(k) \partial_k \mathcal{H}_{\mathbold{\mu}}(k) \right]
  \label{eq:w_def}
\end{equation}
as a 1D topological invariant~\cite{Wen1989641,Sato-PRB-2011}.
The identity with another definition of the winding number, which uses
a flat band Hamiltonian~\cite{Schnyder-PRB-2008}, is proven in Appendix
\ref{subsec:App:Identity_windingNumber}.
Let us consider the unitary transformation $U_\pi$ defined in Eq.~\eqref{eq:U_pi}, 
which rotates the Pauli matrices for the particle-hole basis as
$U_\pi^\dagger \hat{\pi}_x U_\pi = \hat{\pi}_y$,
$U_\pi^\dagger \hat{\pi}_y U_\pi = \hat{\pi}_z$,
$U_\pi^\dagger \hat{\pi}_z U_\pi = \hat{\pi}_x$.
Correspondingly, the Hamiltonian in Eq.~\eqref{eq:H_BdG_bulk_0} takes
in the following an off-diagonal form,
\begin{equation}
  \mathcal{\tilde{H}}_{\mathbold{\mu}}(k) 
  = 
  U_\pi^\dagger \mathcal{H}_{\mathbold{\mu}}(k) U_\pi
  = 
  \left(
    \begin{array}{cc}
      0 & h_{\mathbold{\mu}}(k) \\
      h^\dagger_{\mathbold{\mu}}(k) & 0
    \end{array}
    \right),
  \label{eq:H_BdG_bulk}
\end{equation}
where
\begin{equation}
  h_{\mathbold{\mu}}(k)
  =
  \left(
  \begin{array}{cc}
    \varepsilon_{\mathrm{c},\mathbold{\mu}}(k) - i s \Delta_0 & f_{\mathrm{e},\mathbold{\mu}}(k) - i s f_{\mathrm{eh},\mu}(k) \\
    f_{\mathrm{e},\mathbold{\mu}}^*(k) - i s f_{\mathrm{eh},\mu}^*(k) & \varepsilon_{\mathrm{c},\mathbold{\mu}}(k) - i s \Delta_0
  \end{array}
  \right).
  \label{eq:h_matrix}
\end{equation}
At the same time, the fermionic operators are transformed as
\begin{equation}
  \left(
  \begin{array}{c}
    c_{\chi \sigma  k \mathbold{\mu}} \\
    c_{\xi \sigma k \mathbold{\mu}}
  \end{array}
  \right)
\equiv U_\pi^{-1}
  \left(
  \begin{array}{c}
    c_{\sigma k \mathbold{\mu}} \\
    c_{\sigma -k -\mathbold{\mu}}^\dagger
  \end{array}
  \right).
  \label{eq:c_unitaryTransform}
\end{equation}
Because the chiral operator is transformed as $\tilde{\Gamma} =
U_\pi^\dagger \Gamma U_\pi = \hat{\pi}_z$, the winding number is
written as
\begin{equation}
  w_{\mathbold{\mu}} = \frac{1}{2 \pi} \int_{-\pi/a_z}^{\pi/a_z} dk \partial_k \arg \det h_{\mathbold{\mu}}(k),
  \label{eq:w_int_arg_det_h}
\end{equation}
with the determinant of $h_{\mathbold{\mu}}(k)$ being
\begin{align}
  \det h_{\mathbold{\mu}}
  = & \varepsilon_{\mathrm{c},\mathbold{\mu}}^2 - \Delta_0^2 - |f_{\mathrm{e},\mathbold{\mu}}|^2 + |f_{\mathrm{eh},\mu}|^2 \nonumber \\
  & + 2 i s \left( \varepsilon_{\mathrm{c},\mathbold{\mu}} \Delta_0 + \frac{f_{\mathrm{e},\mathbold{\mu}} f_{\mathrm{eh},\mu}^* + f_{\mathrm{e},\mathbold{\mu}}^* f_{\mathrm{eh},\mu}}{2} \right)
  \label{eq:deth}
\end{align}
(see Appendix~\ref{subsec:App:windingN} for the derivation).

For the case of $|\Delta_0|, |\Delta_1| \ll |\mu_\mathrm{c}|,
|\gamma|$, on which we are focusing, the real part of $\det
h_{\mathbold{\mu}}$ is expressed as
\begin{align}
  \mathrm{Re} \left( \det h_{\mathbold{\mu}} \right) 
  & 
  \simeq \varepsilon_{\mathrm{c},\mathbold{\mu}}^2 - |f_{\mathrm{e},\mathbold{\mu}}|^2 
\nonumber \\
  & 
  = 
  \left[ -\mu_\mathrm{c} + \varepsilon_{\mathrm{so},\mathbold{\mu}} \right]^2 - |f_{\mathrm{e},\mathbold{\mu}}|^2.
\end{align}

\begin{figure}[tb]
  \includegraphics[width=8.5cm]{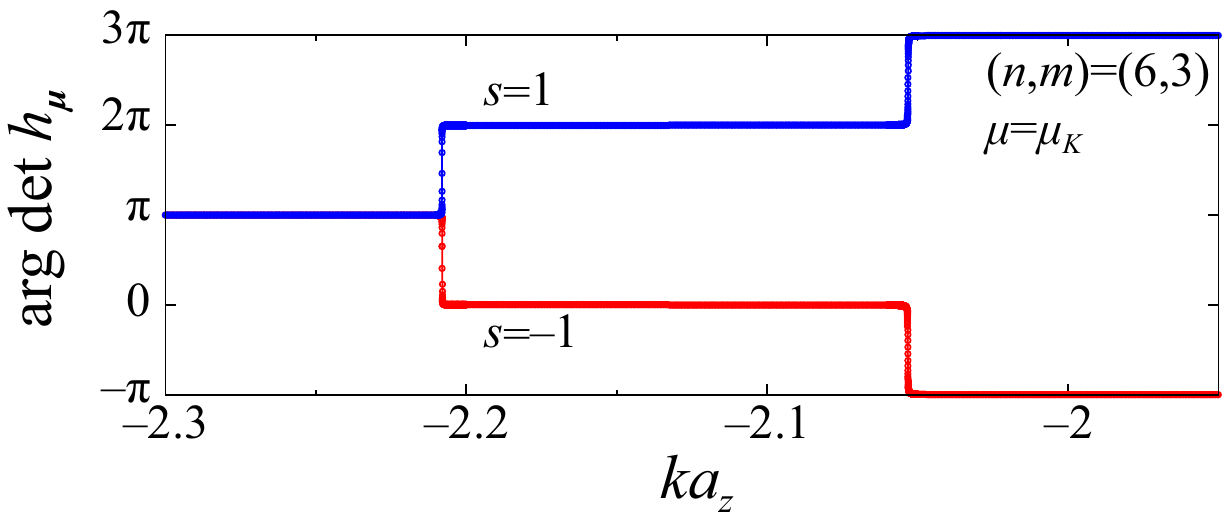}
  \caption{
    Phase of $\det h_{\mathbold{\mu}}$, $\arg \det
    h_{\mathbold{\mu}}$, appearing in the integrand of the winding
    number in Eq.~\eqref{eq:w_int_arg_det_h}, for an $(n,m)=(6,3)$
    nanotube near the $K$ point for which the angular momentum is
    $\mu_K=2$.
    The parameters are the same as those in
    Fig.~\ref{fig:0603_Esck}(b).
    The continuous change of the function in the interval $-\pi \le
    \arg \det h_{\mathbold{\mu}} \le 3 \pi$ is clearly seen.
    The blue and red curves show the spin components $s=1$ and $-1$,
    respectively.
    Note that both of them are almost equal $\pi$ in the regions of $k
    a_z \lesssim -2.21$.
    For this case, the integrand gives contribution $+1$ ($-1$) to the
    winding number for $s=1$ ($s=-1$).
  }
  \label{fig:0603_argdeth_windingN}
\end{figure}

\noindent
Except near the Fermi points, we have
\begin{equation}
  \left| \mathrm{Re} \left( \det h_{\mathbold{\mu}} \right) \right| 
  \gg \left| \mathrm{Im} \left( \det h_{\mathbold{\mu}} \right) \right|
\end{equation}
since the imaginary part of $\det h_{\mathbold{\mu}}$ is
proportional to the superconducting pairing potentials $\Delta_0$ and
$\Delta_1$.
Therefore, Eq.~\eqref{eq:deth} is approximated as a positive or
negative real number, and then the phase of $\det h_{\mathbold{\mu}}$
is almost constant and equal to $0$ or $\pi$.
This feature is clearly observed in
Fig.~\ref{fig:0603_argdeth_windingN}, which shows the phase of the
determinant of $h_{(\mu_K,s)}$ near the $K$ point.

Let us focus on the regions near the Fermi points at the $\tau$ valley,
which are the only ones where the phase of $\det h_{\mathbold{\mu}_\tau}$ changes
and a finite contribution to the integral in
Eq.~\eqref{eq:w_int_arg_det_h} is expected, as can be seen in
Fig.~\ref{fig:0603_argdeth_windingN}.
As seen in the $\mathbold{k}\cdot\mathbold{p}$ scheme, in which the
functions $f_{\mathrm{e},\mathbold{\mu}}$ and
$f_{\mathrm{eh},\mathbold{\mu}}$ have the form in
Eq.~\eqref{eq:f_expansion}, $\mathrm{Re} \left( \det
h_{\mathbold{\mu}_\tau} \right)$ behaves quadratically in $k$ near the
$\tau$ point.
That is, $\mathrm{Re} \left( \det h_{\mathbold{\mu}_\tau} \right)$ is
negative for $k_z < k_-^{(\mathbold{\tau})}$ and $k_z >
k_+^{(\mathbold{\tau})}$, and is positive for $k_-^{(\mathbold{\tau})}
< k_z < k_+^{(\mathbold{\tau})}$.
Note that the two roots of $\mathrm{Re} \left( \det
h_{\mathbold{\mu}_\tau} \right)$ are regarded as the two Fermi points
in our approximation of small superconducting couplings.

\begin{figure}[tb]
  \centering\includegraphics[width=6cm]{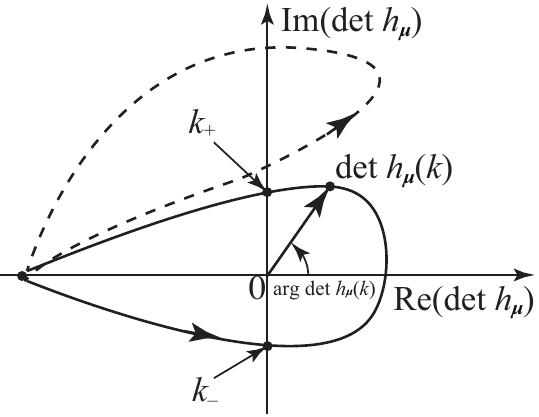}
  \caption{
    Schematics of the trajectories of the complex function $\det
    h_{\mathbold{\mu}}$ in the complex plane when $k$ changes from $k
    \ll k_-$ to $k \gg k_+$.
    The solid curve shows an example for a nontrivial winding number
    $w_\mathbold{\mu}=1$, and the dashed curve shows a case for a
    trivial winding number $w_\mathbold{\mu}=0$.
  }
  \label{fig:windingh}
\end{figure}
%

\begin{figure}[tb]
  \includegraphics[width=8.5cm]{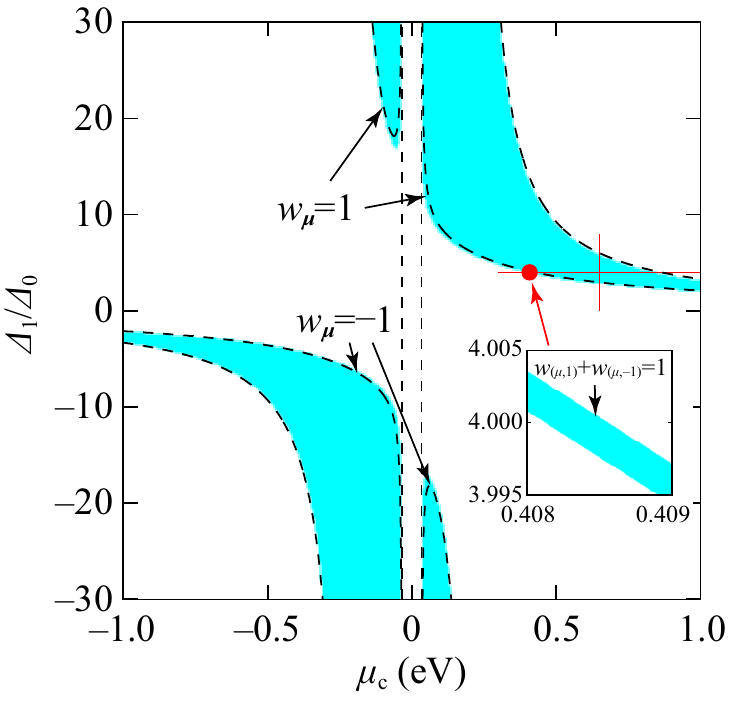}
  \caption{
    Topological phase diagram for an $(n,m)=(6,3)$ nanotube estimated
    from Eq.~\eqref{eq:cond_nontrivial_w} for $(\tau,s)=(1,1)$ in the
    $\mu_\mathrm{c}$ and $\Delta_1 / \Delta_0$ plane, where
    $\Delta_0=0.5$~meV.
    The light blue areas show the region of nontrivial winding number,
    $|w_\mathbold{\mu}|=1$.
    The dashed curves show Eq.~\eqref{eq:TPD_border}, the analytical
    expression for the border of the topological phases.
    The region between the dashed vertical lines is the band gap region
    of the normal state.
    The red lines indicate the parameter region of
    Fig.~\ref{fig:0603_BdG_Finite}.
    The inset shows the phase diagram for the value
    $w_{(\mu,1)}+w_{(\mu,-1)}$ near the region marked by the red
    point, which has a nontrivial value only near the border of the
    main figure.
}
  \label{fig:0603_TPD}
\end{figure}

Let us define $h_{r}^{(\mathbold{\tau})} \equiv
h_{\mathbold{\mu}_\tau}(k_{r}^{(\mathbold{\tau})})$,
the function $h_{\mathbold{\mu}_\tau}$ at the Fermi point for the $\tau$ valley.
When $\mathrm{Im} ( \det h_{+}^{(\mathbold{\tau})} )$ has the opposite
sign of $\mathrm{Im} ( \det h_{-}^{(\mathbold{\tau})} )$,
\begin{equation}
  \mathrm{Im} \left( \det h_{+}^{(\mathbold{\tau})} \right) 
  \mathrm{Im} \left( \det h_{-}^{(\mathbold{\tau})} \right) < 0,
  \label{eq:cond_nontrivial_w}
\end{equation}
then $\det h_{\mathbold{\mu}}$ near the Dirac point contributes to a
nontrivial winding number (see the schematics in
Fig.~\ref{fig:windingh}).
Note that the maximum contribution to the winding number per Dirac
point is $|w_{\mathbold{\mu}}|=1$ because of the above discussion.
The sign of the winding number is given by the sign of $\mathrm{Im} (
\det h_{+}^{(\mathbold{\tau})} )$, that is, the winding number is
\begin{equation}
  w_{\mathbold{\mu}} = \sgn [ \mathrm{Im} ( \det h_{+}^{(\mathbold{\tau})} ) ]
  |w_{\mathbold{\mu}}|.
\end{equation}
Figure~\ref{fig:0603_TPD} shows the topological phase diagram for an
$(n,m)=(6,3)$ nanotube calculated from
Eq.~\eqref{eq:cond_nontrivial_w} for $(\tau,s)=(1,1)$.

Within the $\mathbold{k}\cdot\mathbold{p}$ approximation, after some
algebra given in Appendix~\ref{subsec:App:windingN}, the condition
\eqref{eq:cond_nontrivial_w} is summarized as
\begin{align}
  & \left[
    \frac{\gamma}{\mu_{\mathrm{c},\mathbold{\tau}}}
    + \left( \frac{\Delta_1}{\Delta_0} \right) \left( 1 + \varepsilon_{c,\mathbold{\tau}} E_{c,\mathbold{\tau}} \right)
    \right]^2 \nonumber \\
  & \quad
  - \varepsilon_{z,\mathbold{\tau}}^2 \left( \frac{\Delta_1}{\Delta_0} \right)^2 \left( 1 - E_{c,\mathbold{\tau}}^2 \right) < 0.
  \label{eq:cond_nontrivial_w_expr}
\end{align}
Using the relation
\begin{equation}
  \mathrm{Im} \left( \det h_{r}^{(\mathbold{\tau})} \right)
  = s \mu_{\mathrm{c},\mathbold{\tau}} \varepsilon_{\mathrm{g}, r}^{(\mathbold{\tau})},
  \label{eq:deth_epsg}
\end{equation}
which is given in Appendix~\ref{sec:App:BdGspectrum} [after
  Eq.~\eqref{eq:eps_g_halfway}], the sign of the winding number can
also be evaluated.
As seen in Eq.~\eqref{eq:cond_nontrivial_w_expr}, the condition holds
only when $\varepsilon_{z,\mathbold{\tau}} \neq 0$, that is, $\Delta
k_z \neq 0$, the case of a finite shift of the Dirac point in the
axial direction, and $\Delta_1 \neq 0$.
Note that $E_{c,\mathbold{\tau}}^2 < 1$ holds outside the energy gap
of the nanotubes.
As shown in Eq.~\eqref{eq:app:Deltakc_Deltakz}, we have a finite
$\Delta k_z$ except for the pure zigzag SWNTs, for which the chiral
angle is $\theta = 0$.
We also notice that the condition
\eqref{eq:cond_nontrivial_w_expr} depends on the ratio of
$\Delta_0$ and $\Delta_1$ but not on their absolute values.

At the border, one of the two superconducting gaps
$\varepsilon_{\mathrm{g}, r}^{(\mathbold{\tau})}$ ($r = \pm$)
becomes zero.
Then, from the condition $\varepsilon_{\mathrm{g},
  r}^{(\mathbold{\tau})} = 0$ and Eq.~\eqref{eq:egap}, the border is
determined by
\begin{align}
  & \frac{\Delta_1}{\Delta_0}
  = \nonumber \\
  & - \frac{\gamma}{\mu_{\mathrm{c},\mathbold{\tau}}}
  \frac{
    \left( 1 + \varepsilon_{c,\mathbold{\tau}} E_{c,\mathbold{\tau}} \right)
    + r \tau \sgn(\mu_{\mathrm{c},\mathbold{\tau}}) \sqrt{ \varepsilon_{z,\mathbold{\tau}}^2 \left( 1 - E_{c,\mathbold{\tau}}^2 \right) }
  }
       {
         \left( 1 + \varepsilon_{c,\mathbold{\tau}} E_{c,\mathbold{\tau}} \right)^2 - \varepsilon_{z,\mathbold{\tau}}^2 \left( 1 - E_{c,\mathbold{\tau}}^2 \right)
       }.
       \label{eq:TPD_border}
\end{align}
Note that the border is also given by the roots of the left-hand side
of Eq.~\eqref{eq:cond_nontrivial_w_expr}.
By comparing with the numerical calculation in
Fig.~\ref{fig:0603_BdG_Finite}, we confirm that the zero energy edge
states appear in the region where the winding number has a nonzero
value.
The region becomes narrower and the borders asymptotically behave as
$\Delta_1/\Delta_0 \simeq - \gamma / \mu_\mathrm{c}$ for a large
$\mu_\mathrm{c}$.
This implies that to have nontrivial winding number, the ratio $\delta
= \Delta_1/\Delta_0$ becomes smaller and comparable to 1 for
$|\mu_\mathrm{c}| \sim |\gamma|$, as shown in Fig.~\ref{fig:0603_TPD}.
However, such a chemical potential might be unrealistically large.

Let us comment on the effect of the spin-orbit interaction.
As shown in Eq.~\eqref{eq:Deltak_muc_SO}, the spin-orbit interaction
gives the spin dependence in the phase diagram.
Since we are focusing on the conducting region for the normal state,
we have $|\mu_\mathrm{c}| \gg |\varepsilon_\mathrm{so}|$.
Furthermore, we also have the relation $|\Delta k_c| \gg |\Delta
k_\mathrm{so}|$ except for the armchair SWNTs.
For the armchair SWNTs, the spin-orbit interaction opens a small gap
at $\mu_\mathrm{c} = 0$, as already pointed out in previous
studies~\cite{Ando-2000-06,Chico-2004-10,Huertas-Hernando-2006-10,Izumida-2009-06}.
Therefore, we have an almost identical phase diagram for the $(\tau,s)$
and $(\tau,-s)$ subspaces except for the sign difference reflecting
the opposite winding direction between $s$ and $-s$, as seen in the
relation \eqref{eq:deth_epsg}.
A small difference between the opposite spins, shown as the finite
value of $w_{(\mu_K,1)}+w_{(\mu_K,-1)}$ in the inset of
Fig.~\ref{fig:0603_TPD}, appears at the border region scaled by the
spin-orbit interaction.
Note that the phase diagrams for $(\tau,s)$ and $(-\tau,-s)$ are the
same including the sign.
Therefore, the total winding number, $\sum_{\tau, s}
w_{\mathbold{\mu}_\tau} = 2 (w_{(\mu_K,1)}+w_{(\mu_K,-1)})$, shows the
same diagram as $w_{(\mu_K,1)}+w_{(\mu_K,-1)}$.
As a result, the \textit{total} winding number is nonzero only in very
narrow regions of the parameter space.
Nevertheless, several edge states are present in the nanotube even
when the total winding number is zero, which proves that the total
topological invariant may miss a rich part of the physics of the
system.

As a further example, it should be noted that in the armchair class,
with $\mu_k = 0 = \mu_{K'}$, the winding number $w_\mathbold{\mu}$
becomes zero even when the condition
\eqref{eq:cond_nontrivial_w_expr} is satisfied for both valleys.
This is because the winding directions for the $\tau$ and $-\tau$
valley are opposite, which can be seen from the relation
Eq.~\eqref{eq:deth_epsg}.
However, this does not mean that there is no edge state for the
armchair class, as will be discussed in Sec.~\ref{sec:armchairClass}.

Let us comment on the symmetry class to which our 1D model belongs
according to the topological classification in
Ref.~\cite{Schnyder-PRB-2008}.
Since we have only the chiral symmetry in each $\mathbold{\mu}$
subspace, the 1D model in that space belongs to the AIII class.
The total Hamiltonian has time-reversal symmetry, and belongs to class
DIII.
Further discussion on the different topological invariants in our
system can be found in Appendix~\ref{subsec:App:otherTIs}.

It should be noted that the nontrivial topological phase obtained in
our work does not contradict a previous study~\cite{Haim-PRB-2016},
which predicts only a trivial topological phase if the induced
superconducting correlation is $s$ wave.
This correlation appears in our Hamiltonian as the onsite pairing.
As already mentioned, the $\Delta_1$ term, which is the coupling
constant for the $k$-linear term in Eq.~\eqref{eq:f_expansion}, and
thus acts as the $p$-wave superconducting
coupling~\cite{Uchoa-PRL-2007}, is needed to have the nontrivial
topological phases.

\section{Bulk-edge correspondence}
\label{sec:Bulk-edge}

In this section we shall reveal the deep physical meaning of the
condition constituting Eq.~\eqref{eq:cond_nontrivial_w}.
As mentioned in the Introduction, it has been
shown~\cite{Izumida-2016-05} for the SWNTs in the normal state that
the winding number per $\mathbold{\mu}$ space $w_{\mathbold{\mu}}$ is
equal to the number of edge states in this space.
The latter are given by the difference between the number of
evanescent modes, being the solutions of the mode equation at zero
energy, and the number of boundary conditions for given sublattice.
This gives a one-to-one correspondence between the winding number as a
topological invariant and the physical edge state.
This kind of relation is called a bulk-edge correspondence.
Let us discuss the bulk-edge correspondence for the present system by
including the finite length of the SWNT in our description.

Since the relevant contribution to the winding number comes from the neighborhood of 
the $\tau$ point, we shall consider an effective 1D continuum model
obtained by expanding around the $\tau$ point.
The envelope function,
\begin{equation}
  \Psi_{\mathbold{\tau}}
  = 
  \left(
  \begin{array}{c}
    \Psi_{\chi \mathbold{\tau}} \\
    \Psi_{\xi \mathbold{\tau}}
  \end{array}
  \right),
\quad
  \Psi_{p \mathbold{\tau}} = 
  \left(
  \begin{array}{c}
    \Psi_{p \mathrm{A} \mathbold{\tau}} \\
    \Psi_{p \mathrm{B} \mathbold{\tau}}
  \end{array}
  \right),
\end{equation}
obeys the equation,
\begin{align}
  \hat{\mathcal{\tilde{H}}}_{\mathbold{\mu}_\tau}(\hat{k}_z) \Psi_{\mathbold{\tau}} 
  = \varepsilon \Psi_{\mathbold{\tau}},
\end{align}
where $p = \chi, \xi$, and
$\hat{\mathcal{\tilde{H}}}_{\mathbold{\mu}_\tau}(\hat{k}_z)$ has the
same functional form of Eq.~\eqref{eq:H_BdG_bulk} with
Eq.~\eqref{eq:f_expansion}.
However, the wave number $k_z$ is now regarded as the operator
\begin{equation}
  \hat{k}_z = - i \frac{\partial}{\partial z}
\end{equation}
in the continuum model.
At zero energy, $\varepsilon = 0$, the equation can be divided into
two sets of equations with $2 \times 2$ matrix forms:
\begin{equation}
  \hat{h}_{p \mathbold{\mu}_\tau}(\hat{k}_z) \Psi_{p \mathbold{\tau}} = 0,
\end{equation}
where 
$\hat{h}_{\chi \mathbold{\mu}_\tau}(\hat{k}_z)$ and 
$\hat{h}_{\xi \mathbold{\mu}_\tau}(\hat{k}_z)$ are given by 
changing $k_z \rightarrow \hat{k}_z$ 
in 
$h_{\mathbold{\mu}_\tau}^\dagger(k_z)$ and $h_{\mathbold{\mu}_\tau}(k_z)$, respectively.
Let us consider the modes with the following form:
\begin{equation}
  \Psi_{p \mathbold{\tau}} 
  = 
  e^{i q z}
  \left(
  \begin{array}{c}
    1 \\
    \eta_{p}
  \end{array}
  \right).
  \label{eq:Psi_relations}
\end{equation}

In each $p$ block, the modes obey the following equation:
\begin{widetext}
\begin{equation}
  \left(
  \begin{array}{l}
    - \mu_{\mathrm{c},\mathbold{\tau}} + i p s \Delta_0 \quad\quad\quad\quad c \left[ \gamma \left( q - \tau \Delta k_z \right) + i p s \Delta_1 q \right] + c \left[ i \gamma \left( k_c - \Delta k_{c,\mathbold{\tau}} \right) - p s \Delta_1 k_c \right] \\
    c^* \left[ \gamma \left( q - \tau \Delta k_z \right) + i p s \Delta_1 q \right] - c^* \left[ i \gamma \left( k_c - \Delta k_{c,\mathbold{\tau}} \right) - p s \Delta_1 k_c \right] \quad\quad\quad\quad - \mu_{\mathrm{c},\mathbold{\tau}} + i p s \Delta_0
  \end{array}
  \right)
  \left(
  \begin{array}{c}
    1 \\
    \eta_{p}
  \end{array}
  \right)
  = 0,
  \label{eq:eq_qmode}
\end{equation}
\end{widetext}
where we have alternatively used the index $p = 1$ and $-1$ for $p
= \chi$ and $\xi$, respectively.
To have nontrivial solutions, the determinant of the matrix in
Eq.~\eqref{eq:eq_qmode} should be zero.
Since this gives a second-order equation in $q$,
there exist two modes corresponding to the solutions $q_{r}^{(\mathbold{\tau})}$.
A relation between $q_{r}^{(\mathbold{\tau})}$ and the Fermi point
$k_{r}^{(\mathbold{\tau})}$ will be shown in Eq.~\eqref{eq:q_r_sols}.

Within the continuum model, the microscopic boundary condition is
implicitly taken into account in order to form eigenstates.
They are constructed as linear combinations of two independent modes,
a leftgoing and a rightgoing one, subject to boundary conditions at each
end.
Note that in the superconducting gap region the two modes are two
decaying modes, that is, $|\kappa_{r}^{(\mathbold{\tau})}| < 1$ or
$|\kappa_{r}^{(\mathbold{\tau})}| > 1$, where $\kappa_r \equiv
\mathrm{Im} ( q_{r}^{(\mathbold{\tau})} )$.
If the two modes have the same decaying direction, 
that is,
\begin{equation}
  \kappa_{+}^{(\mathbold{\tau})} \kappa_{-}^{(\mathbold{\tau})} > 0,
  \label{eq:cond_edgeState}
\end{equation}
then an edge state given by the linear combination of the two modes
appears at an end.
In the following we explicitly show that this condition is identical
to the condition \eqref{eq:cond_nontrivial_w} for nontrivial
winding number.

As shown in Appendix~\ref{sec:App:Anal_1D_cont}, we arrive after some
algebra to the two solutions
\begin{equation}
  \mathrm{Re} \left( q_{r}^{(\mathbold{\tau})} \right) \simeq k_{r}^{(\mathbold{\tau})}, 
  \quad
  \kappa_{r}^{(\mathbold{\tau})} \simeq 
  \frac{r p s \sgn(\mu_{\mathrm{c},\mathbold{\tau}})}{|c| \gamma \sqrt{1 - E_{c,\mathbold{\tau}}^2}}
  \frac{\varepsilon_{\mathrm{g}, r}^{(\mathbold{\tau})}}{2}.
  \label{eq:q_r_sols}
\end{equation}
Since we have the relation \eqref{eq:deth_epsg}, we get
\begin{equation}
  \kappa_{r}^{(\mathbold{\tau})}
  =
  \frac{r p ~ \mathrm{Im} \left( \det h_{r}^{(\mathbold{\tau})} \right)}{4 |\mu_{\mathrm{c},\mathbold{\tau}}| |c| \gamma \sqrt{1 - E_{c,\mathbold{\tau}}^2}}.
  \label{eq:kappa_Imdeth}
\end{equation}
Combining Eqs.~\eqref{eq:cond_edgeState} and \eqref{eq:kappa_Imdeth},
it is immediately clear that the condition for emergence of an edge
state is identical to the condition for a nontrivial winding number
expressed by Eq.~\eqref{eq:cond_nontrivial_w}.

It is worth noting that, from Eq.~\eqref{eq:q_r_sols}, the
decay length of the edge state is proportional to the Fermi velocity,
$-|c| \gamma \sqrt{1 - E_{c,\mathbold{\tau}}^2} / \hbar$, of the normal states at the
given chemical potential and is inversely proportional to the
superconducting gap.
This implies the shortest decay length to be near the bottom of the
conduction or top of the valence bands for the semiconducting SWNTs.

\section{Armchair class}
\label{sec:armchairClass}

So far, we have been restricting ourselves to the case of decoupled
valleys.
Let us discuss the effect of valley coupling by considering the
armchair class SWNTs, in which the two valleys have the same angular
momentum.

\begin{figure}[tb]
  \includegraphics[width=8.5cm]{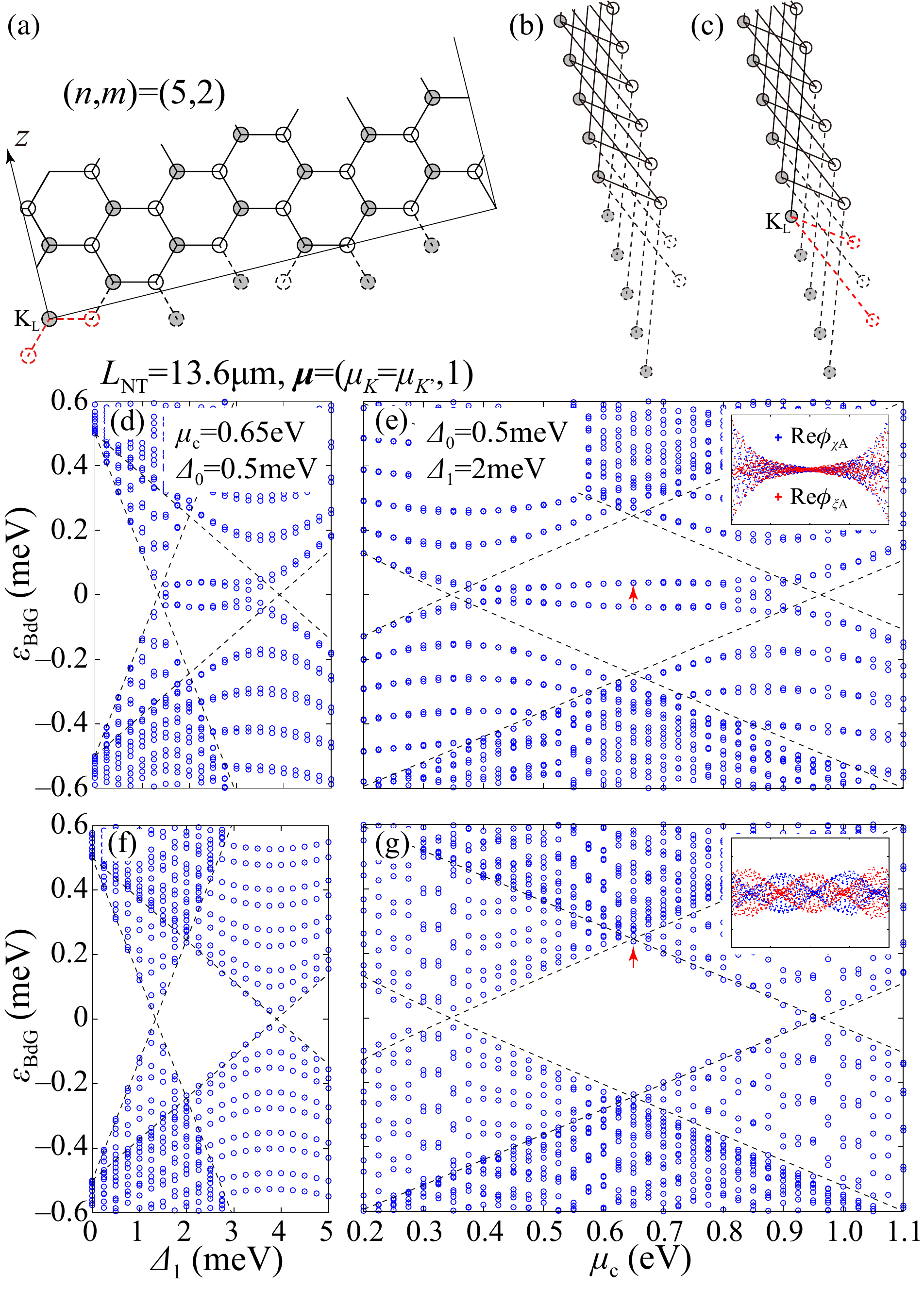}
  \caption{
    BdG spectrum of an armchair class $(5,2)$ nanotube with length of
    $13.6$~$\mu$m in the $\mathbold{\mu}=(\mu,s)=(\mu_K,+1)$ subspace.
    (a) Unrolled tube near the left end.
    The boundary is formed by a simple cut of the lattice in the plane
    orthogonal to the nanotube axis, represented by the solid line
    perpendicular to the $z$ axis.
    Removed lattice sites adjacent to the boundary sites are
    represented by dashed circles, and the dangling bonds are
    represented by the dashed lines.
    The orthogonal boundary is given by keeping the Klein site
    indicated by $\mathrm{K_L}$, and the minimal boundary is given by
    removing the Klein site.
    (b) Minimal and (c) orthogonal boundaries, respectively, in the 1D
    model.
    (d) BdG spectrum as a function of the superconducting pairing
    $\Delta_1$, and, (e) that as a function of the chemical potential
    $\mu_\mathrm{c}$, respectively, for the minimal boundary, and, (f)
    and (g) show those for the orthogonal boundary.
    Each inset in (e) and (g) shows the real part components
    $\phi_{\chi \mathrm{A}}$ (blue) and $\phi_{\xi \mathrm{A}}$ (red)
    in arbitrary units as a function of lattice site $\ell$ for
    the calculated eigenfunction at the state indicated by the red
    arrow.
}
  \label{fig:0502_BdG_Finite}
\end{figure}

In previous studies~\cite{Izumida-2015-06,Marganska-PRB-2015}, it has
been shown that 
the nature of the valley coupling depends on the boundary conditions.
Here we consider two types of boundaries.
One is the minimal boundary, in which the edge has minimum number of
dangling bonds [see Figs.~\ref{fig:0502_BdG_Finite}(a) and
\ref{fig:0502_BdG_Finite}(b)].
Another is the orthogonal boundary formed by a simple cut of the
lattice in the plane orthogonal to the nanotube axis [see
Figs.~\ref{fig:0502_BdG_Finite}(a) and \ref{fig:0502_BdG_Finite}(c)].
The two valleys are nearly decoupled for the former case, while they
strongly couple for the latter case,
where each eigenstate is formed from a leftgoing mode at one
valley and a rightgoing mode at another
valley~\cite{Izumida-2015-06}.

Figure~\ref{fig:0502_BdG_Finite} shows the calculated spectrum for an
$(n,m)=(5,2)$ nanotube with $N_L = 4 \times 10^5$, which corresponds
to the nanotube length of $13.6$~$\mu$m, in the subspace of
$\mathbold{\mu} = (\mu_K, 1)$, where $\mu_K=\mu_{K'}=0$.
Figures~\ref{fig:0502_BdG_Finite}(d) and \ref{fig:0502_BdG_Finite}(e), which show the case of
the minimal boundary, exhibit a spectrum similar to that in
Fig.~\ref{fig:0603_BdG_Finite}.
Edge states near zero energy are seen inside the gap region for $1
\lesssim \Delta_1 \lesssim 4$~meV in Fig.~\ref{fig:0502_BdG_Finite}(d), and for $350 \lesssim \mu_c
\lesssim 950$~meV in Fig.~\ref{fig:0502_BdG_Finite}(e).
A small deviation from zero energy is observed because of weak valley
coupling.
On the other hand, Figs.~\ref{fig:0502_BdG_Finite}(f) and
\ref{fig:0502_BdG_Finite}(g), which show the case of the orthogonal
boundary, do not support zero energy states in the same region of
superconducting pairing and chemical potential as in
Figs.~\ref{fig:0502_BdG_Finite}(d) and \ref{fig:0502_BdG_Finite}(e).
This is in contrast to the zigzag class SWNTs, in which the shape of
the boundary does not affect the number of edge states if $\mu$
remains a good quantum number since the two valleys have different
$\mu$ and they are decoupled.

The absence of zero energy states for the case of strong valley
coupling in Figs.~\ref{fig:0502_BdG_Finite}(f) and
\ref{fig:0502_BdG_Finite}(g) can be captured by the expressions we
have obtained in Sec.~\ref{sec:Bulk-edge}.
Between the two states specified by $(\tau,s,r)$ and $(-\tau,s,-r)$,
which form a pair for an eigenstate under the boundary condition, we
always have the relation $\kappa_{r}^{(\tau,s)} \simeq
-\kappa_{-r}^{(-\tau,s)}$ because $\varepsilon_{\mathrm{g},
  r}^{(\tau,s)} \simeq \varepsilon_{\mathrm{g}, r}^{(-\tau,s)}$.
Therefore, the condition \eqref{eq:cond_edgeState} of emergence of
edge states is never satisfied for this case.

\section{Discussion and Conclusion}
\label{sec:Conclusion}

We have studied the edge states in the proximity-induced
superconducting gap of finite-length SWNTs from the topological
viewpoint.
Our analysis shows that the numerically observed
edge states are due to the combined effect of curvature-induced Dirac
point shifting and strong superconducting coupling between nearest-neighbor
sites.
A 1D continuum model reveals that the condition for nontrivial
winding number coincides with the condition for emergence of edge
states in the finite length case.

We have seen that in our setup the edge states of zigzag and armchair
classes with the minimal boundary are formed not by time-reversal
symmetric partners, but by the $(\tau,s,r)$ and $(\tau,s,-r)$ states.
Here, $\tau$ is the index of the two valleys $K$ and $K'$, $s$ is that
of spin direction $\uparrow$ and $\downarrow$, and $r$ is that of
left and right branch of the energy bands.
In armchair class with the orthogonal boundary it was impossible to
construct an edge mode, because that required combining $(\tau,s,r)$
and $(-\tau,s,-r)$ states, which always decay in the opposite
directions.

The zero energy edge states studied in this paper appear in pairs
because of the unbroken time-reversal symmetry as well as the
decoupling of two valleys.
As seen in Fig. \ref{fig:0502_BdG_Finite}, mixings of subspaces such
as spin mixing induced, e.g., by an external magnetic field or valley
mixing induced, e.g., by broken rotational symmetry would couple the two
pair members and they would deviate from the zero energy.
These properties would be in contrast to those of the Majorana bound
states, which emerge under breaking of the time-reversal symmetry and
might further necessitate valley mixing, as shown, e.g., in the previous
study~\cite{Sau-PRB-2013}.
Therefore, the control of the magnetic field as well as the rotational
symmetry provides us with a tool for discriminating between the zero
energy edge states discussed in this paper and the Majorana bound
states.

Finally, it is interesting to comment on the possibility of Majorana
bound states in the SWNTs.
If the parameters of the system can be tuned in such a way that the
bound states combine two time-reversal partners $(\tau,s,r)$ and
$(-\tau,-s,-r)$, the requirement of the same decay direction
$\kappa_r^{(\boldsymbol{\tau})} \kappa_{-r}^{(-\boldsymbol{\tau})} >
0$ follows automatically from Eq.~\eqref{eq:q_r_sols}.
This can be achieved in the presence of the spin mixing and the valley
mixing induced by, e.g., an external magnetic field and a potential
scattering.
Furthermore, fine tuning of the system parameters under the spin-orbit
splitting could provide an odd number of time-reversal symmetric
pairs.
As discussed in Refs.~\cite{Egger-PRB-2012,Sau-PRB-2013}, the
combined time-reversal symmetric partners may form edge states of
Majorana nature.

\begin{acknowledgments}
  We acknowledge JSPS KAKENHI Grants (No. JP15K05118, No. JP15KK0147,
  and No. JP16H01046), Grant No. GRK 1570, IGK Topological Insulators.
  W.I. is grateful to A. Yamakage and R. Okuyama for fruitful
  discussion.
  M.M. would like to thank K. Flensberg and M. Wimmer for helpful
  remarks.
\end{acknowledgments}

\appendix

\section{Tight-binding Hamiltonian}
\label{sec:App:TBH}

Here we show the details of the tight-binding Hamiltonian.

\subsection{Tight-binding Hamiltonian of the normal term}
\label{subsec:App:TBH_normal}

Here, we give the explicit form of the vectors
$\mathbold{\delta}_j^{(t)}$ and the hopping integrals
$\gamma_{s,j}^{(t)}$ which appear in the standard term $H_0$ of the
Hamiltonian in Eq.~\eqref{eq:H_0}:
\begin{align}
  H_0
  = 
  & - \mu_\mathrm{c}
  \sum_{\mathbold{r} \sigma s}
  c_{\sigma \mathbold{r} s}^\dagger c_{\sigma \mathbold{r} s}  \nonumber \\
  & + \sum_{\mathbold{r} s} 
  \sum_{j=1}^3 \gamma_{s,j}^{(1)}
  c_{\mathrm{A} \mathbold{r} s}^\dagger c_{\mathrm{B} \mathbold{r} + \mathbold{\delta}_j^{(1)} s} + \mathrm{H.c.} 
  \nonumber \\
  & + \sum_{\mathbold{r} \sigma s} 
  \sum_{j=1}^3 \gamma_{s,j}^{(2)}
  c_{\sigma \mathbold{r} s}^\dagger c_{\sigma \mathbold{r} + \mathbold{\delta}_j^{(2)} s}
  + \mathrm{H.c.}
\end{align}
The vectors to the three nearest-neighbor B sites from the A site are given by
$\mathbold{\delta}_1^{(1)} = \left( \mathbold{a}_1 + \mathbold{a}_2 \right) / 3$,
$\mathbold{\delta}_2^{(1)} = \left( \mathbold{a}_1 - 2 \mathbold{a}_2 \right) / 3$,
and $\mathbold{\delta}_3^{(1)} = \left( -2 \mathbold{a}_1 + \mathbold{a}_2 \right) / 3$.
The vectors to the six next-nearest-neighbor sites are given by
$\mathbold{\delta}_1^{(2)} = \mathbold{a}_1$,
$\mathbold{\delta}_2^{(2)} = \left( \mathbold{a}_1 - \mathbold{a}_2 \right)$,
$\mathbold{\delta}_3^{(2)} = - \mathbold{a}_2$,
and 
$\mathbold{\delta}_{j}^{(2)} = - \mathbold{\delta}_{j-3}^{(2)}$ for $j=4,5,6$.
The hopping integrals are expressed as~\cite{Okuyama-2017-01}
\begin{align}
  \gamma_{s,j}^{(1)}
  & = \gamma \left[ 1 + \Delta k_c \frac{a}{\sqrt{3}} \sin \phi_j 
  - \left( \Delta k_{z} + i s \Delta k_\mathrm{so} \right) \frac{a}{\sqrt{3}} \cos \phi_j
  \right], \\
  \gamma_{s,j}^{(2)}
  & = i \frac{\left( -1 \right)^{j+1}}{3 \sqrt{3}} s \varepsilon_\mathrm{so},
\end{align}
where $\gamma = -2.57$~eV, $\phi_j = \theta - 5 \pi / 6 + 2 \pi j /
3$.
Finally, $\theta = \arccos \frac{2n + m}{2 \sqrt{n^2 + m^2 + nm}}$ is
the chiral angle defined as the angle between $\mathbold{C}_h$ and
$\mathbold{a}_1$, and $d_t = \left| \mathbold{C}_h \right| / \pi = a
\sqrt{n^2 + m^2 + nm} / \pi$ is the diameter of nanotube.
The terms proportional to 
\begin{equation}
  \Delta k_z = \zeta \frac{\sin 3 \theta}{d_t^2}, \quad
  \Delta k_c = \beta' \frac{\cos 3 \theta}{d_t^2},
  \label{eq:app:Deltakc_Deltakz}
\end{equation}
with $\zeta = -0.185$~nm, $\beta = 0.0436$~nm account for the
curvature-induced shift of the Dirac point from the $K$ point in the
axial and the circumferential directions, respectively.
The pure imaginary hopping terms proportional to
\begin{equation}
  \Delta k_\mathrm{so} = \alpha_1' V_\mathrm{so} \frac{1}{d_t}, \quad
  \varepsilon_\mathrm{so} = \alpha_2 V_\mathrm{so} \frac{\cos 3 \theta}{d_t},
  \label{eq:app:Deltakso_epsso}
\end{equation}
represent the spin-dependent Dirac point shift in the circumferential
direction and the Zeeman-like energy shift, respectively, induced by
the spin-orbit interaction.
We use $\alpha_1' = 8.8 \times 10^{-5}$~meV$^{-1}$, $\alpha_2 =
-0.045$~nm and $V_\mathrm{so} = 6$~meV.
All numerical values above, except for $V_\mathrm{so}$, have been
obtained by fitting to the results of \textit{ab initio} based
tight-binding calculation in Ref.~\cite{Izumida-2009-06}.
Note that the relation 
$( \gamma_{-s,j}^{(t)} )^* = \gamma_{s,j}^{(t)}$ ($t = 1,2$) 
holds reflecting the time-reversal symmetry.

\subsection{Transformation of the $\Delta_1$ term to the BdG form}
\label{subsec:App:BdG_Delta1}

Here we show the transformation of the superconducting term
proportional to $\Delta_1$ in Eq.~\eqref{eq:H_s_1Dlattice} to the BdG
form in Eq.~\eqref{eq:H_BdG_1Dlattice}.
The operators can be doubled such as
\begin{align}
& \sum_{\mathbold{\mu}} 
  e^{i \frac{2 \pi}{d} \delta \nu_j^{(1)} \mu}
  s c_{\mathrm{A} \ell \mathbold{\mu}}^\dagger
  c_{\mathrm{B} \ell_j^{(1)} -\mathbold{\mu}}^\dagger \nonumber \\
& = \frac{1}{2}
  \sum_{\mathbold{\mu}}
  e^{i \frac{2 \pi}{d} \delta \nu_j^{(1)} \mu}
  s 
  \left( 
  c_{\mathrm{A} \ell \mathbold{\mu}}^\dagger
  c_{\mathrm{B} \ell_j^{(1)} -\mathbold{\mu}}^\dagger 
  -
  c_{\mathrm{B} \ell_j^{(1)} -\mathbold{\mu}}^\dagger 
  c_{\mathrm{A} \ell \mathbold{\mu}}^\dagger
  \right) 
  \nonumber \\
& = \frac{1}{2}
  \sum_{\mathbold{\mu}}
  s 
  \left( 
  e^{i \frac{2 \pi}{d} \delta \nu_j^{(1)} \mu}
  c_{\mathrm{A} \ell \mathbold{\mu}}^\dagger
  c_{\mathrm{B} \ell_j^{(1)} -\mathbold{\mu}}^\dagger 
  \right. \nonumber \\
& \quad
  \quad
  \quad
  \quad
  \quad
  \left.
  +
  e^{-i \frac{2 \pi}{d} \delta \nu_j^{(1)} \mu}
  c_{\mathrm{B} \ell_j^{(1)} \mathbold{\mu}}^\dagger 
  c_{\mathrm{A} \ell -\mathbold{\mu}}^\dagger
  \right).
\end{align}
From the second to the third equations, we have exchanged
$\mathbold{\mu} \leftrightarrow -\mathbold{\mu}$ for the second term.
Then, the contribution proportional to $\Delta_1$ in
Eq.~\eqref{eq:H_s_1Dlattice} is rewritten as
\begin{align}
& \Delta_1 
  \sum_{\mathbold{\mu}} 
  e^{i \frac{2 \pi}{d} \delta \nu_j^{(1)} \mu}
  s c_{\mathrm{A} \ell \mathbold{\mu}}^\dagger
  c_{\mathrm{B} \ell_j^{(1)} -\mathbold{\mu}}^\dagger + \mathrm{H.c.} \nonumber \\
= & 
  \frac{\Delta_1}{2}
  \sum_{\mathbold{\mu}} s 
  \left( 
  e^{i \frac{2 \pi}{d} \delta \nu_j^{(1)} \mu}
  c_{\mathrm{A} \ell \mathbold{\mu}}^\dagger
  c_{\mathrm{B} \ell_j^{(1)} -\mathbold{\mu}}^\dagger
  \right. \nonumber \\
& 
  \quad
  \quad
  \quad
  \quad
  \left.
  +
  e^{-i \frac{2 \pi}{d} \delta \nu_j^{(1)} \mu}
  c_{\mathrm{B} \ell_j^{(1)} \mathbold{\mu}}^\dagger
  c_{\mathrm{A} \ell -\mathbold{\mu}}^\dagger
  \right)
 + \mathrm{H.c.} \nonumber \\
= & 
  \frac{\Delta_1}{2}
  \sum_{\mathbold{\mu}} s 
  e^{i \frac{2 \pi}{d} \delta \nu_j^{(1)} \mu}
  \left( 
  c_{\mathrm{A} \ell \mathbold{\mu}}^\dagger
  c_{\mathrm{B} \ell_j^{(1)} -\mathbold{\mu}}^\dagger
  +
  c_{\mathrm{A} \ell -\mathbold{\mu}}
  c_{\mathrm{B} \ell_j^{(1)} \mathbold{\mu}}
  \right)
 + \mathrm{H.c.} \nonumber \\
= & 
  \frac{\Delta_1}{2}
  \sum_{\mathbold{\mu}} s 
  e^{i \frac{2 \pi}{d} \delta \nu_j^{(1)} \mu}
  \mathbf{c}_{\mathrm{A} \ell \mathbold{\mu}}^\dagger
  \hat{\pi}_x
  \mathbf{c}_{\mathrm{B} \ell_j^{(1)} \mathbold{\mu}}
 + \mathrm{H.c.},
\end{align}
where $\mathbf{c}_{\sigma \ell \mathbold{\mu}}^\dagger$,
$\mathbf{c}_{\sigma \ell \mathbold{\mu}}$ are the Nambu spinors
introduced in Eq.~\eqref{eq:NambuSpinor}.
From the second to third equations we have exchanged the term
$\Delta_1 e^{-i \frac{2 \pi}{d} \delta \nu_j^{(1)} \mu}
  c_{\mathrm{B} \ell_j^{(1)} \mathbold{\mu}}^\dagger
  c_{\mathrm{A} \ell -\mathbold{\mu}}^\dagger$
and its Hermite conjugate
$\Delta_1 e^{i \frac{2 \pi}{d} \delta \nu_j^{(1)} \mu}
  c_{\mathrm{A} \ell -\mathbold{\mu}}
  c_{\mathrm{B} \ell_j^{(1)} \mathbold{\mu}}$.

\section{BdG excitation spectrum of the bulk system}
\label{sec:App:BdGspectrum}

In this appendix, we show the detailed analytical calculation of the
BdG excitation spectrum near the Dirac points.
Such spectrum is obtained by diagonalizing the Hamiltonian matrix of
Eq.~\eqref{eq:H_BdG_bulk_0}, or, equivalently,
Eq.~\eqref{eq:H_BdG_bulk}.
Hereafter, we omit the subscripts $\mathbold{\mu}$, $\mu$ and $k$, for
simplicity.

It follows from the chiral symmetry $\{\tilde{\Gamma}, \mathcal{\tilde{H}}\}=0$
that for the eigenfunction $\Psi$ satisfying $\mathcal{\tilde{H}} \Psi = \varepsilon \Psi$
there exists a paired state
$\tilde{\Gamma} \Psi$ with the energy $-\varepsilon$, that is, 
$\mathcal{\tilde{H}} \tilde{\Gamma} \Psi = - \varepsilon \tilde{\Gamma} \Psi$.
And, multiplying the Schr\"odinger equation by the Hamiltonian, one gets
$\mathcal{\tilde{H}}^2 \Psi = \varepsilon^2 \Psi$.
Since $\mathcal{\tilde{H}}$ has a block-off-diagonal form [see
Eq.~\eqref{eq:H_BdG_bulk}], $\mathcal{\tilde{H}}^2$ has the block-diagonal form with blocks $h h^\dagger$ and $h^\dagger h$.
Then two separated equations, $h h^\dagger \Psi_\chi = \varepsilon^2
\Psi_\chi$ and $h^\dagger h \Psi_\xi = \varepsilon^2 \Psi_\xi$, are
obtained, where $\Psi = ( \Psi_\chi, \Psi_\xi )^T$.
The eigenvalue problem can be reduced to solving the equation $\det
\left[ h h^\dagger - \varepsilon^2 I_{2 \times 2} \right] = 0$, which
gives,
\begin{align}
  & \varepsilon^4
  - 2 \left( \varepsilon_\mathrm{c}^2 + \Delta_0^2 + |f_\mathrm{e}|^2 + |f_\mathrm{eh}|^2 \right) \varepsilon^2 
  \nonumber \\
  & + 
  \left[ \left( \varepsilon_\mathrm{c}^2 - \Delta_0^2 \right) 
      - \left( |f_\mathrm{e}|^2 - |f_\mathrm{eh}|^2 \right) \right]^2
  \nonumber \\
  & + 
  \left[ 
    2 \varepsilon_\mathrm{c} \Delta_0 
    + \left( f_\mathrm{e} f_\mathrm{eh}^* + f_\mathrm{e}^* f_\mathrm{eh} \right)
    \right]^2 = 0.
  \label{eq:app:BdGeigenvalueEq}
\end{align}
The equation $\det \left[ h^\dagger h - \varepsilon^2 I_{2 \times 2}
  \right] = 0$ is the same as Eq.~\eqref{eq:app:BdGeigenvalueEq}.
Since the eigenvalue equation is quadratic in $\varepsilon^2$, we have
\begin{equation}
  \varepsilon^2 = A \pm 2 \sqrt{B},
\end{equation}
with
\begin{align}
  A = & \varepsilon_\mathrm{c}^2 + \Delta_0^2 + |f_\mathrm{e}|^2 + |f_\mathrm{eh}|^2, \\
  B = & \left( \Delta_0^2 + |f_\mathrm{e}|^2 \right) \left( \varepsilon_\mathrm{c}^2 + |f_\mathrm{eh}|^2 \right) 
  \nonumber \\
  & - \left( \frac{f_\mathrm{e} f_\mathrm{eh}^* + f_\mathrm{e}^* f_\mathrm{eh}}{2} 
  + \varepsilon_\mathrm{c} \Delta_0 \right)^2.
\end{align}
Then, the BdG spectrum is given by
\begin{equation}
  \varepsilon = \pm \sqrt{A \pm 2 \sqrt{B}}.
\end{equation}

Let us focus on the BdG spectrum near the superconducting gap by
expanding $\varepsilon = \pm \sqrt{A - 2 \sqrt{B}}$ near the Fermi
points by using Eq.~\eqref{eq:f_expansion}.
Near the two Fermi points $k_z = k_r$ ($r = \pm$), which are given in
Eq.~\eqref{eq:k_r}, it holds
$\varepsilon_\mathrm{c} = - \mu_{\mathrm{c},\mathbold{\tau}}$, and
\begin{equation}
  f_\mathrm{e} = f_{\mathrm{e} r} + c \gamma k', 
  \quad
  f_\mathrm{eh} \simeq f_{\mathrm{eh} r} = \frac{\Delta_1}{\gamma} f_{\mathrm{e} r} + C,
  \label{eq:app:f_e_f_eh_expansion}
\end{equation}
where $k'$ is the 1D wave number measured from $k_z = k_r$.
Furthermore, $f_{\mathrm{e} r}$ and $f_{\mathrm{eh} r}$ are $f_{\mathrm{e}}$ and
$f_{\mathrm{eh}}$ at the Fermi point, respectively,
\begin{align}
  & f_{\mathrm{e} r} = c \gamma \left[ \left( k_r - \tau \Delta k_z \right) + i \left( k_c - \Delta k_{c,\mathbold{\tau}} \right) \right], 
  \label{eq:app:f_er} \\ 
  & C = c \Delta_1 \left( \tau \Delta k_z + i \Delta k_{c,\mathbold{\tau}} \right),
  \label{eq:app:C}
\end{align}
and we have discarded the weak $k'$ dependence of $f_\mathrm{eh}$.
Note that the relation $|f_{\mathrm{e} r}| = |\mu_{\mathrm{c},\mathbold{\tau}}|$
holds.
$A$ and $B$ are expanded near each Fermi point as
\begin{align}
  A & = A_0 + A_1 k' + A_2 k'^2 + \cdots, \\
  B & = B_0 + B_1 k' + B_2 k'^2 + \cdots,
\end{align}
where
\begin{align}
  A_0 & = 2 \mu_{\mathrm{c},\mathbold{\tau}}^2 + \Delta_0^2 + |f_{\mathrm{eh} r}|^2, \\
  A_1 & = \gamma \left( c^* f_{\mathrm{e} r} + c f_{\mathrm{e} r}^* \right),
\end{align}
and
\begin{align}
  B_0 = & \mu_{\mathrm{c},\mathbold{\tau}}^4
  + \mu_{\mathrm{c},\mathbold{\tau}}^2 |f_{\mathrm{eh} r}|^2 
  -\frac{1}{4} \left( f_{\mathrm{e} r} f_{\mathrm{eh} r}^* + f_{\mathrm{e} r}^* f_{\mathrm{eh} r} \right)^2 
  \nonumber \\
  & - \left( f_{\mathrm{e} r} f_{\mathrm{eh} r}^* + f_{\mathrm{e} r}^* f_{\mathrm{eh} r} \right) \mu_{\mathrm{c},\mathbold{\tau}} \Delta_0, \\
  B_1 = & \mu_{\mathrm{c},\mathbold{\tau}}^2 \gamma \left( c^* f_{\mathrm{e} r} + c f_{\mathrm{e} r}^* \right)
  + c \gamma \left( c^* f_{\mathrm{e} r} + c f_{\mathrm{e} r}^* \right) |f_{\mathrm{eh} r}|^2 \nonumber \\
  & - \frac{1}{2} \left( f_{\mathrm{e} r} f_{\mathrm{eh} r}^* + f_{\mathrm{e} r}^* f_{\mathrm{eh} r} \right)
  \gamma \left( c^* f_{\mathrm{eh} r} + c f_{\mathrm{eh} r}^* \right)
  \nonumber \\
  & - \gamma \left( c^* f_{\mathrm{eh} r} + c f_{\mathrm{eh} r}^* \right) \mu_{\mathrm{c},\mathbold{\tau}} \Delta_0.
\end{align}

Near the gap region, we have
\begin{align}
  \varepsilon^2 
  & = A - 2 \sqrt{B} \nonumber \\
  & = \left( A_0 - 2 \sqrt{B_0} \right)
  + \left( A_1 - \frac{B_1}{\sqrt{B_0}} \right) k' + \cdots.
  \label{eq:app:epsilon2_mid}
\end{align}
Using 
\begin{align}
  \sqrt{B_0} 
  \simeq &
  \mu_{\mathrm{c},\mathbold{\tau}}^2
  \left[
    1 
    + \frac{1}{2} 
    \left( \frac{|f_{\mathrm{eh} r}|^2}{\mu_{\mathrm{c},\mathbold{\tau}}^2} 
    - \frac{ \left( f_{\mathrm{e} r} f_{\mathrm{eh} r}^* + f_{\mathrm{e} r}^* f_{\mathrm{eh} r} \right)^2 }{4 \mu_{\mathrm{c},\mathbold{\tau}}^4} \right. \right. \nonumber \\
    & \left. \left.
    - \frac{ \left( f_{\mathrm{e} r} f_{\mathrm{eh} r}^* + f_{\mathrm{e} r}^* f_{\mathrm{eh} r} \right) \Delta_0 }{\mu_{\mathrm{c},\mathbold{\tau}}^3}
    \right)
    \right],
\end{align}
each coefficient in Eq.~\eqref{eq:app:epsilon2_mid} is expressed as
\begin{equation}
  A_0 - 2 \sqrt{B_0} = \left( \frac{\varepsilon_{\mathrm{g}, r}^{(\mathbold{\tau})}}{2} \right)^2, 
  \quad
  A_1 - \frac{B_1}{\sqrt{B_0}} \simeq 0,
\end{equation}
where 
\begin{equation}
  \varepsilon_{\mathrm{g}, r}^{(\mathbold{\tau})} \equiv 
  \frac{f_{\mathrm{e} r} f_{\mathrm{eh} r}^* + f_{\mathrm{e} r}^* f_{\mathrm{eh} r}}{\mu_{\mathrm{c},\mathbold{\tau}}} + 2 \Delta_0,
  \label{eq:eps_g_halfway}
\end{equation}
and we have discarded the higher order of
$\Delta_{0/1}/\mu_{\mathrm{c},\mathbold{\tau}}$ and $\Delta_{0/1}/\gamma$ in each
contribution.
The coefficient of $k'$ in Eq.~\eqref{eq:app:epsilon2_mid} being zero
means that the gap position is at $k_z = k_{r}^{(\mathbold{\tau})}$ within
this approximation.
Therefore, $\varepsilon_{\mathrm{g}, r}^{(\mathbold{\tau})}$ represents the
superconducting gap at $k_z = k_{r}^{(\mathbold{\tau})}$.
By comparing Eq.~\eqref{eq:deth} and Eq.~\eqref{eq:eps_g_halfway},
we get Eq.~\eqref{eq:deth_epsg}.
By using Eqs.~\eqref{eq:app:f_e_f_eh_expansion}--\eqref{eq:app:C},
we finally get the expression of the superconducting gap given in
Eq.~\eqref{eq:egap}.

\section{Topological invariants}
\label{sec:App:topologicalInvariants}

In this appendix, we discuss properties of the winding number and other
topological invariants.

\subsection{Identity of two expressions for the winding number}
\label{subsec:App:Identity_windingNumber}

Here we show the identity of two different expressions for the
winding number; one is given by Eq.~\eqref{eq:w_def} and another is
given using a flat-band Hamiltonian approach~\cite{Schnyder-PRB-2008}.

First, we shall get a flat-band Hamiltonian.
Let us consider the filled and empty states $\Psi_{\pm, l}(k)$ of the
Hamiltonian \eqref{eq:H_BdG_bulk} with the eigenvalue $\pm
\varepsilon_l(k)$, where $-$ ($+$) refers to the filled (empty) state
and $\varepsilon_l(k) >0$.
The eigenfunctions are written as~\cite{Schnyder-PRB-2012}
\begin{equation}
\Psi_{\pm, l}(k)
=
  \frac{1}{\sqrt{2}} 
  \left(
  \begin{array}{c}
    u_l \\
    \pm \frac{1}{\varepsilon_l(k)} h^\dagger(k) u_l
  \end{array}
  \right),
\end{equation}
where $l$ is the index of the spectrum and we have omitted the
subscript $\mathbold{\mu}$ for simplicity.
One can check that the function satisfies the Schr\"odinger equation 
$\mathcal{\tilde{H}}(k) \Psi_l(k) = \pm \varepsilon_l(k) \Psi_l(k)$ with the
relation $h(k) h^\dagger(k) u_l = \varepsilon_l^2(k) u_l$, which is
derived by multiplying the Schr\"odinger equation by the Hamiltonian.
\textit{A priori} the eigenvectors $u_l$ may depend on $k$.
Nevertheless, since $u_l$ are eigenvectors of $h(k)h^\dag(k)$ which is
Hermitian, they form an orthogonal set.
We can perform a unitary transformation into a basis in which
$hh^\dag$ is diagonal, and the eigenvectors $u_l$ are independent of
$k$.
This transformation is continuous in $k$, which is assured by the
continuity of the original eigenvectors $u_l(k)$.
In the following, we shall work implicitly in that transformed basis.
The projector onto the filled states is given by
\begin{equation}
  P = \sum_l \Psi_{-, l}(k) \Psi_{-, l}(k)^\dagger
  = \frac{1}{2} I - \frac{1}{2} Q.
\end{equation}
The operator $Q$ acts as a flat-band Hamiltonian
having the energy $+1$ for the empty states and $-1$ for the filled states
independent of $k$
since $Q \Psi_{\pm, l}(k) = (I - 2P) \Psi_{\pm, l}(k) = \pm \Psi_{\pm, l}(k)$.
In the matrix form, we have
\begin{equation}
  Q =
  \left(
  \begin{array}{cc}
    0 & q(k) \\
    q^\dagger(k) & 0
  \end{array}
  \right),
\end{equation}
where
\begin{equation}
  q(k) = \sum_l \frac{1}{\varepsilon_l(k)} u_l u_l^\dagger h(k) = U(k) h(k),
\end{equation}
and the matrix
\begin{equation}
  U(k) = \sum_l \frac{1}{\varepsilon_l(k)} u_l u_l^\dagger
  \label{eq:app:U_def}
\end{equation}
has been introduced.
Using the flat-band Hamiltonian, a winding number is defined
as~\cite{Schnyder-PRB-2008}
\begin{equation}
  w'
  = \frac{1}{2 \pi i} 
  \int dk 
  \mathrm{Tr} \left[ q^{-1}(k) \partial_k q(k) \right],
  \label{eq:app:w_def_flatBand}
\end{equation}
where the integral is taken over the whole of the 1D BZ.

Before showing the identity of the two different definitions of the
winding number, let us show a relation which will be used later.
Hereafter, we will also omit $k$ in the expressions for simplicity.
Since $Q^2 = (I - 2P)^2 = I - 4P + 4 P^2 = I$, we have $q q^{\dagger}
= 1$.
Then, $q^{-1} = q^\dagger$.
We also have $U^\dagger = U$, which is immediately seen from
Eq.~\eqref{eq:app:U_def}.
From these two relations we have $h^{-1} U^{-1} = h^\dagger U^\dagger
= h^\dagger U$.
Then, it holds
\begin{equation}
  U^{-1} = h h^\dagger U 
  = \sum_l \varepsilon_l u_l u_l^\dagger.
  \label{eq:app:Uinv}
\end{equation}

Let us show the identity of $w'$ in Eq.~\eqref{eq:app:w_def_flatBand}
with $w$ in Eq.~\eqref{eq:w_def}.
Since $0 = \partial_k \left( q^{-1} q \right) = \left(
\partial_k q^{-1} \right) q + q^{-1} \partial_k q$, the integrand in
the expression of the winding number $w'$ is written as
\begin{equation}
  \mathrm{Tr} \left[ q^{-1} \partial_k q \right]
  = 
  \frac{1}{2} \mathrm{Tr} \left[ q^{-1} \partial_k q - q \partial_k q^{-1} \right].
\end{equation}
By using the relations $\partial_k q = \left( \partial_k U \right) h +
U \partial_k h$, $\partial_k q^{-1} = \left( \partial_k h^{-1} \right)
U^{-1} + h^{-1} \partial_k U^{-1}$, and the cyclic property of the
trace, we have
\begin{align}
  & \mathrm{Tr} \left[ q^{-1} \partial_k q - q \partial_k q^{-1} \right] \nonumber \\
  & = \mathrm{Tr} \left[ \left( h^{-1} \partial_k h - h \partial_k h^{-1} \right)
    + \left( U^{-1} \partial_k U - U \partial_k U^{-1} \right) \right].
\end{align}
The second term is rewritten as
\begin{align}
  & \mathrm{Tr} \left[ U^{-1} \partial_k U - U \partial_k U^{-1} \right]  \nonumber \\
  & = \mathrm{Tr} \left[ U^{-1} \partial_k U - \partial_k \left( U U^{-1} \right) 
    + \left( \partial_k U \right) U^{-1} \right] \nonumber \\
  & = 2 \mathrm{Tr} \left[ U^{-1} \partial_k U \right].
\end{align}
By using Eq.~\eqref{eq:app:Uinv}, 
the term is further calculated as
\begin{align}
  \mathrm{Tr} \left[ U^{-1} \partial_k U \right]
  & = \mathrm{Tr} 
  \left[ \sum_{l l'} \left( - \varepsilon_l \frac{\partial_k \varepsilon_{l'}}{\varepsilon_{l'}^2} \right)
    u_l u_l^\dagger u_{l'} u_{l'}^\dagger \right] \nonumber \\
  & = - \mathrm{Tr} 
  \left[ \sum_l \frac{\partial_k \varepsilon_l}{\varepsilon_l}
    u_l u_{l}^\dagger \right] 
  = - \sum_l \partial_k \log \varepsilon_l,
\end{align}
where we have used the orthogonality $u_l^\dagger u_{l'} =
\delta_{l,l'}$ and $\mathrm{Tr} [ u_l u_{l}^\dagger ] = \sum_m u_{l,m}
u_{l,m}^* = 1$.
Using the periodicity of $\varepsilon_l(k)$ in the BZ, we finally get
the identity
\begin{align}
  w'
  & = \frac{1}{4 \pi i} 
  \int dk 
  \mathrm{Tr} \left[ h^{-1} \partial_k h - h \partial_k h^{-1} \right] \nonumber \\
  & = -\frac{1}{4 \pi i} \int dk \mathrm{Tr} \left[ \tilde{\Gamma} \mathcal{\tilde{H}}^{-1} \partial_k \mathcal{\tilde{H}} \right]
  \nonumber \\
  & = w.
\end{align}

\subsection{Winding number near the Dirac points}
\label{subsec:App:windingN}

Let us show the analytical calculation of the winding number near the
Dirac points.
First we show how to derive the expression
\eqref{eq:w_int_arg_det_h} from Eq.~\eqref{eq:w_def}.
Using the unitary matrix \eqref{eq:U_pi} and the transformed
Hamiltonian \eqref{eq:H_BdG_bulk}, the winding number is expressed
as
\begin{align}
  w_{\mathbold{\mu}} 
  & = -\frac{1}{4 \pi i} \int dk \mathrm{Tr} \left[ \tilde{\Gamma} \mathcal{\tilde{H}}_{\mathbold{\mu}}^{-1} \partial_k \mathcal{\tilde{H}}_{\mathbold{\mu}} \right]
  \nonumber \\
  & = \frac{1}{4 \pi i} \int dk \mathrm{Tr} \left[ h_{\mathbold{\mu}}^{-1} \partial_k h_{\mathbold{\mu}} - h_{\mathbold{\mu}}^{\dagger -1} \partial_k h_{\mathbold{\mu}}^\dagger \right] 
  \nonumber \\
  & = \frac{1}{4 \pi i} \int dk \left( \partial_k \log \det h_{\mathbold{\mu}} - \partial_k \log \det h_{\mathbold{\mu}}^\dagger \right) 
  \nonumber \\
  & = \frac{1}{2 \pi} \mathrm{Im} \int dk \partial_k \log \det h_{\mathbold{\mu}} \nonumber \\
  & = \frac{1}{2 \pi} \int dk \partial_k \arg \det h_{\mathbold{\mu}},
\end{align}
where 
we have used the formulas
\begin{align}
  & \mathrm{Tr} \left[ h^{-1} \partial_k h \right] = \partial_k \log \det h, \\
  & \log \det h^\dagger = \mathrm{Re} \left( \log \det h \right) - i \mathrm{Im} \left( \log \det h \right).
\end{align}

Next, we will show the calculation leading from
Eq.~\eqref{eq:cond_nontrivial_w} to
Eq.~\eqref{eq:cond_nontrivial_w_expr}.
Near the $\tau$ point the imaginary part of $\det
h_{\mathbold{\mu}_\tau}$ is given by
\begin{align}
  \frac{\mathrm{Im} \left( \det h_{\mathbold{\mu}_\tau} \right)}{2 s}
  = & 
  \mu_{\mathrm{c},\mathbold{\tau}} \Delta_0 + |c|^2 \gamma \Delta_1 
  \left[ k_z \left( k_z - \tau \Delta k_z \right)
    \right. \nonumber \\
    & 
    \left. + k_c \left( k_c - \Delta k_{c,\mathbold{\tau}} \right) 
    \right].
\end{align}
At the Fermi points $k_z = k_{r}^{(\mathbold{\tau})}$ the imaginary part
is calculated as
\begin{align}
  \frac{\mathrm{Im} \left( \det h_{r}^{(\mathbold{\tau})} \right)}{2 s} \frac{\gamma}{\mu_{\mathrm{c},\mathbold{\tau}}^2 \Delta_0} 
  = & \frac{\gamma}{\mu_{\mathrm{c},\mathbold{\tau}}}
  + \frac{\Delta_1}{\Delta_0}
  \left( 1 + \varepsilon_{c,\mathbold{\tau}} E_{c,\mathbold{\tau}} \right)
  \nonumber \\
  - & r \frac{\Delta_1}{\Delta_0}
  \tau \varepsilon_{z,\mathbold{\tau}} \sgn( \mu_{\mathrm{c},\mathbold{\tau}} ) \sqrt{ 1 - E_{c,\mathbold{\tau}}^2}.
\end{align}
Then, the condition \eqref{eq:cond_nontrivial_w} is summarized in
the form of Eq.~\eqref{eq:cond_nontrivial_w_expr}.

\subsection{Relation between $\mathbb{Z}$ and $\mathbb{Z}_2$ invariants}
\label{subsec:App:otherTIs}

We have shown that an integer ($\mathbb{Z}$) topological invariant,
the winding number, can be defined for our system.
The periodic table of topological
invariants~\cite{Chiu-RevModPhys-2016} nevertheless states that a DIII
class Hamiltonian has a paritylike $\mathbb{Z}_2$ invariant.
These two facts appear contradictory, but are not, as we will now
clarify.
The discussion is based on the approach to topological invariants
presented in the review by Chiu \textit{et
  al}~\cite{Chiu-RevModPhys-2016}.

The fundamental topological invariant in 1D is Zak's
phase~\cite{Zak-PRL-1989} in one band carrying a generic index $l$,
\begin{equation}
  \gamma_l = \frac{i}{2\pi}\int_{\mathrm{BZ}} dk \;\mathcal{A}_l(k),
\end{equation}
where $\mathcal{A}_l(k)$ is the Berry connection in band $l$,
$\mathcal{A}_l(k) = \langle \Psi_l(k) \vert \partial_k \vert \Psi_l(k) \rangle$, 
and $\vert \Psi_l(k) \rangle$ is the eigenfunction of a 1D bulk
Hamiltonian for eigenvalue $\varepsilon_l$.
Since the Berry connection is gauge dependent, so is Zak's phase,
but it can be shown that a gauge transformation changes $\gamma_l$
only by an integer.
The more frequently used invariant is therefore $W = \exp \left (2 \pi
i \sum_l \gamma_l \right)$, where $l$ are the indices of filled bands,
which is gauge independent, although in general not quantized.
The presence of discrete symmetries restricts the values which
$\gamma_l$ can take.
In systems with chiral symmetry, in the gauge given by the chiral
basis the winding number can be shown to be $\mathbb{Z} \ni w_l = 2
\gamma_l$, therefore $W = \exp(i\pi \sum_{l} w_{l}) = \pm 1$.
In systems with particle-hole symmetry the topological invariant $W$
can be evaluated using the representation of the Hamiltonian in the
Majorana basis,
\begin{equation}
  \mathcal{H}(k) = U_{\mathrm{M}}^\dagger [i X(k)] U_{\mathrm{M}}.
\end{equation}
At time reversal invariant momenta $k=0, \pi$, $X(k)$ is real and
skew-symmetric, $X(k) = -[X(k)]^T$.
The topological invariant $W$ can then be expressed through the
Pfaffian of $X$,
\begin{equation}
  W = \mathrm{sgn} \left\{ \mathrm{Pf} [X(\pi)]\;\mathrm{Pf}[X(0)] \right\}
  = \pm 1,
\end{equation}
which is of a $\mathbb{Z}_2$ type.
In our system the complete $16 \times 16$ Hamiltonian $\mathcal{H}(k)$
has the four-block-diagonal form,
\begin{align}
  & \mathcal{H}(k) = \nonumber \\
  & \mathrm{diag}
  \left[ 
    H_{(\mu_K,\uparrow)}(k), H_{(\mu_{K'},\downarrow)}(k),
    H_{(\mu_K,\downarrow)}(k), H_{(\mu_{K'},\uparrow)}(k)
    \right],
\end{align}
where $H_{\boldsymbol{\mu}}$ are defined in
Eq.~\eqref{eq:H_BdG_bulk_0}.
The transformation $U_{\mathrm{M}}$, which brings the system into the
Majorana basis, acts in the upper two and the lower two blocks
separately, given as $U_{\mathrm{M}} = \mathrm{diag} \left[
  U_{\mathrm{M}}', U_{\mathrm{M}}' \right]$, where
\begin{equation}
  U_{\mathrm{M}}' = \frac{1}{\sqrt{2}}
  \left(
  \begin{array}{cccccccc}
    1 & 0 & 0 & 0 & 0 & 0 &  1 & 0 \\
    i & 0 & 0 & 0 & 0 & 0 & -i & 0 \\
    0 & 1 & 0 & 0 & 0 & 0 & 0 &  1 \\
    0 & i & 0 & 0 & 0 & 0 & 0 & -i \\
    0 & 0 & 1 & 0 & 1  & 0 & 0 & 0 \\
    0 & 0 & i & 0 & -i & 0 & 0 & 0 \\
    0 & 0 & 0 & 1 & 0 &  1 & 0 & 0 \\
    0 & 0 & 0 & i & 0 & -i & 0 & 0
  \end{array}
  \right),
\end{equation}
resulting in
\begin{equation}
  X(k) 
  = 
  \mathrm{diag}
  \left[
    X_{(\mu_K,\uparrow),(\mu_{K'},\downarrow)}(k), X_{(\mu_K,\downarrow),(\mu_{K'},\uparrow)}
    \right].
\end{equation}
The Pfaffian at $k = 0, \pi$ is given by
\begin{equation}
  \mathrm{Pf}[X(k)] 
  = 
  \mathrm{Pf}[X_{(\mu_K,\uparrow),(\mu_{K'},\downarrow)}(k)]\;\mathrm{Pf}[X_{(\mu_K,\downarrow),(\mu_{K'},\uparrow)}(k)],
\end{equation}
where
\begin{align}
  \mathrm{Pf} [ X_{\mathbold{\mu},-\mathbold{\mu}} ]
  = &
  \left[ \left( \varepsilon_{\mathrm{c},\mathbold{\mu}}^2 - \Delta_0^2 \right) 
    - \left( |f_{\mathrm{e},\mathbold{\mu}}|^2 - |f_{\mathrm{eh},\mathbold{\mu}}|^2 \right) \right]^2
  \nonumber \\
& + 
  \left[ 
    2 \varepsilon_{\mathrm{c},\mathbold{\mu}} \Delta_0 
    + \left( f_{\mathrm{e},\mathbold{\mu}} f_{\mathrm{eh},\mathbold{\mu}}^* 
    + f_{\mathrm{e},\mathbold{\mu}}^* f_{\mathrm{eh},\mathbold{\mu}} \right)
    \right]^2.
\end{align}
Since the Pfaffian is always non-negative, the topological invariant
$W$ is also trivial.
Indeed, our total winding number is always even, $w =
\sum_{\boldsymbol{\mu}} w_{\boldsymbol{\mu}} = 0,\pm 2$, therefore, the
corresponding $W = +1$.
Our nanotube from the $\mathbb{Z}_2$ point of view is always in the
trivial phase.
Nevertheless, a total invariant does not give the full information
about the system.
It is especially clear in quantum spin Hall insulators, where the
total Chern number, summed over two spin directions, vanishes but the
edge states exist for both spins and even are topologically
protected~\cite{Sheng-PRL-2006}.
The information carried by the partial invariants is therefore more
useful.

As a last remark, in contrast to the topological insulators, the edge
states generated by the four $\boldsymbol{\mu}$ subspaces of our
system are not topologically protected, as can be seen from
Figs.~\ref{fig:0502_BdG_Finite}(d) and \ref{fig:0502_BdG_Finite}(e),
where the valley mixing clearly gaps them. Our system is then more
similar to a weak topological insulator, where the states generated by
the nontrivial weak partial invariant can be gapped by disorder,
i.e., a breaking of translational invariance~\cite{Fu-PRL-2007}.

\begin{widetext}

\section{Analysis of the 1D continuum model}
\label{sec:App:Anal_1D_cont}

In this appendix, we will show the detailed calculation of the modes of
the 1D continuum model introduced in Sec.~\ref{sec:Bulk-edge}.
The condition that the determinant of the matrix in
Eq.~\eqref{eq:eq_qmode} is zero is written as
\begin{equation}
  \left( \mu_{\mathrm{c},\mathbold{\tau}} - i p s \Delta_0 \right)^2 
  - |c|^2 \left[ \gamma \left( q - \tau \Delta k_z \right) + i p s \Delta_1 q \right]^2 
  + |c|^2 \left[ i \gamma \left( k_c - \Delta k_c \right) - p s \Delta_1 k_c \right]^2 = 0. 
  \label{eq:deth_curv}
\end{equation}
This is a second-order equation in $q$, and the two
solutions are given by,
\begin{align}
  q_{\mp}^{(\mathbold{\tau})} 
  = & 
  \frac{1}{|c| \left( \gamma + i p s \Delta_1 \right)} 
  \left[
    \pm 
    \sqrt{ \left( \mu_{\mathrm{c},\mathbold{\tau}} - i p s \Delta_0 \right)^2 
      - |c|^2 \left[ \gamma \left( k_c - \Delta k_c \right) + i p s \Delta_1 k_c \right]^2 } 
    + |c| \gamma \tau \Delta k_z
    \right] 
  \nonumber \\
  \simeq & \frac{\gamma - i p s \Delta_1}{|c| \gamma^2}
  \left[
    \pm \sqrt{ \mu_{\mathrm{c},\mathbold{\tau}}^2 - \left[ |c| \gamma \left( k_c - \Delta k_c \right) \right]^2 } 
    \sqrt{ 1 - 2 i p s F}
    + |c| \gamma \tau \Delta k_z
    \right],
\end{align}
where the signs $-$ and $+$ in the index of
$q_{\mp}^{(\mathbold{\tau})}$ correspond to the signs of $+$ and $-$
in the right-hand side, respectively, and
\begin{equation}
  F = 
  \frac{1}{\mu_{\mathrm{c},\mathbold{\tau}}}
  \frac{\Delta_0 + \Delta_1 E_{c,\mathbold{\tau}} \left( E_{c,\mathbold{\tau}} + \varepsilon_{c,\mathbold{\tau}} \right) \frac{\mu_{\mathrm{c},\mathbold{\tau}}}{\gamma}}
       {1 - E_{c,\mathbold{\tau}}^2}.
\end{equation}
Note that $F$ is of the order of $\Delta_{0/1}/\mu_{\mathrm{c},\mathbold{\tau}}$.
By using the formula
\begin{equation}
  \sqrt{a + i b} = \sqrt{ \frac{a + \sqrt{a^2 + b^2} }{2} } 
  + i \sgn(b) \sqrt{ \frac{-a + \sqrt{a^2 + b^2} }{2} }, 
\end{equation}
for $a, b \in \mathbb{R}$, the term $\sqrt{ 1 - 2 i p s F}$ becomes,
\begin{equation}
  \sqrt{ 1 - 2 i p s F} = R + i I,
\end{equation}
where $R$ and $I$ are given by
\begin{align}
  R & = \sqrt{ \frac{1 + \sqrt{1 + (2 p s F)^2} }{2} } 
  \simeq \sqrt{ \frac{1 + 1 + \frac{1}{2} (2 p s F)^2 }{2} } \simeq 1,
  \nonumber \\
  I & = \sgn(-2psF) \sqrt{ \frac{-1 + \sqrt{1 + (2 p s F)^2} }{2} } 
  \nonumber \\
  & \simeq -ps \sgn(F) \sqrt{ \frac{-1 + 1 + \frac{1}{2} (2 p s F)^2 }{2} }
  = -ps F.
\end{align}
Then, for the real part of $q_{\mathbold{\tau},\pm}$, we get
\begin{equation}
  \mathrm{Re} \left( q_{\mathbold{\tau},\pm} \right)
  = 
  \mp 
  \frac{\gamma R + p s \Delta_1 I}{|c| \gamma^2} 
  \sqrt{ \mu_{\mathrm{c},\mathbold{\tau}}^2 - \left[ |c| \gamma \left( k_c - \Delta k_c \right) \right]^2 } 
  + \tau \Delta k_z 
  \simeq 
  \pm \sqrt{ \left( \frac{\mu_{\mathrm{c},\mathbold{\tau}}}{|c| \gamma} \right)^2 - \left( k_c - \Delta k_c \right)^2 } 
  + \tau \Delta k_z 
= k_{\pm}.
\end{equation}
On the other hand, for the imaginary part of
$q_{\mathbold{\tau},\pm}$,
\begin{equation}
  \mathrm{Im} \left( q_{\mathbold{\tau},\pm} \right)
  =
    \frac{1}{|c| \gamma} 
    \left(
    \mp \frac{\gamma I - p s \Delta_1 R}{\gamma} |\mu_{\mathrm{c},\mathbold{\tau}}| \sqrt{ 1 - E_{c,\mathbold{\tau}}^2 } 
    - p s \tau \frac{\Delta_1}{\gamma} |c| \gamma \Delta k_z 
    \right),
\end{equation}
and the numerator of the first term in the right-hand side is
calculated as,
\begin{equation}
  \gamma I - p s \Delta_1 R
  = 
  \frac{-ps}{1 - E_{c,\mathbold{\tau}}^2}
  \left[
    \frac{\gamma}{\mu_{\mathrm{c},\mathbold{\tau}}} \Delta_0 
    + \left( 1 + E_{c,\mathbold{\tau}} \varepsilon_{c,\mathbold{\tau}} \right) \Delta_1 
    \right].
\end{equation}
Then, we get
\begin{align}
  \mathrm{Im} \left( q_{\mathbold{\tau},{\pm}} \right)
  & =
  \pm
  \frac{ps}{|c| \gamma}
  \left\{
  \frac{1}{\sqrt{1 - E_{c,\mathbold{\tau}}^2}}
  \left[
    \frac{\gamma}{\mu_{\mathrm{c},\mathbold{\tau}}} \Delta_0 + \left( 1 + E_{c,\mathbold{\tau}} \varepsilon_{c,\mathbold{\tau}} \right) \Delta_1 
    \right]
  \frac{|\mu_{\mathrm{c},\mathbold{\tau}}|}{\gamma} 
  \mp \tau \Delta_1 \varepsilon_{z,\mathbold{\tau}} \frac{\mu_{\mathrm{c},\mathbold{\tau}}}{\gamma}
  \right\} \nonumber \\
  & = \pm \frac{ps}{|c| \gamma}
  \frac{\sgn(\mu_{\mathrm{c},\mathbold{\tau}})}{\sqrt{1 - E_{c,\mathbold{\tau}}^2}}
  \left[
    \Delta_0 + \Delta_1 \frac{\mu_{\mathrm{c},\mathbold{\tau}}}{\gamma} 
    \left( 1 + E_{c,\mathbold{\tau}} \varepsilon_{c,\mathbold{\tau}} 
    \mp \sgn(\mu_{\mathrm{c},\mathbold{\tau}}) \tau \varepsilon_{z,\mathbold{\tau}} \sqrt{1 - E_{c,\mathbold{\tau}}^2}
    \right)
    \right] \nonumber \\
  & = \pm \frac{ps}{|c| \gamma}
  \frac{\sgn(\mu_{\mathrm{c},\mathbold{\tau}})}{\sqrt{1 - E_{c,\mathbold{\tau}}^2}}
  \frac{\varepsilon_{g,\pm}^{(\mathbold{\tau})}}{2},
\end{align}
which is the expression given in Eq.~\eqref{eq:q_r_sols}.
\end{widetext}

\end{document}